\documentclass[letterpaper,11pt]{article}

\usepackage{amsmath,amssymb,bbm}
\usepackage{cite}
\usepackage{bbold}
\pdfoutput=1
\usepackage{graphicx}

\def\Dslash{\hspace{3pt}\raisebox{1pt}{$\slash$} \hspace{-9pt} D}

\begin{document}

\baselineskip=18pt


\begin{flushright}
\end{flushright}
\vspace{.3in}
\vspace{2.0cm}

\begin{center}
{\Large \bf
Light Top Partners for a Light Composite Higgs
}
\\[6pt]
\vspace{1.0cm}
{\bf Oleksii Matsedonskyi$^{a}$, Giuliano Panico$^{b}$ and Andrea Wulzer$^{a,b}$}

\vspace{0.5cm}

\centerline{$^{a}${\it Dipartimento di Fisica e Astronomia and INFN, Sezione di Padova, }}
\centerline{{\it Via Marzolo 8, I-35131 Padova, Italy}}
\centerline{$^{b}${\it Institute for Theoretical Physics, ETH Zurich,
8093 Zurich, Switzerland}}

\end{center}
\vspace{.8cm}

\begin{abstract}
\medskip
\noindent
Anomalously light fermionic partners of the top quark often appear in explicit constructions,
such as the 5d holographic models, where the Higgs is a light composite pseudo Nambu-Goldstone
boson and its potential is generated radiatively by top quark loops.
We show that this is due to a \emph{structural} correlation among the mass of the partners and the one
of the Higgs boson. Because of this correlation, the presence of light partners
could be essential to obtain a realistic Higgs mass.

We quantitatively confirm this generic prediction, which applies to a broad class of composite Higgs models,
by studying the simplest calculable  framework with a composite Higgs, the Discrete Composite Higgs Model. In this
setup we show analytically that the requirement of a light enough Higgs strongly constraints the fermionic spectrum
and makes the light partners appear.

The light top partners thus provide the most promising manifestation
of the composite Higgs scenario at the LHC. Conversely, the lack of observation
of these states can put strong restrictions on the parameter space
of the model. A simple analysis of the $7$-TeV LHC searches presently available
already gives some non-trivial constraint. The strongest bound comes from the exclusion
of the $5/3$-charged partner. Even if no dedicated LHC search exists for this particle,
a bound of $611$~GeV is derived by adapting the CMS search of bottom-like
states in same-sign dileptons.

\end{abstract}

\newpage


\section{Introduction}

The hints of a light Higgs boson have never been so strong. On top of the indirect indications
coming from the ElectroWeak Precision Tests (EWPT) of LEP we now have powerful direct
constraints from the LHC searches of ATLAS \cite{ATLAS:2012ae} and CMS \cite{Chatrchyan:2012tx}.
As of today, a SM-like Higgs boson is constrained in a very narrow interval around $120$~GeV. \footnote{
The heavy Higgs region, above $\sim500$~GeV, is not covered by the direct searches and could become
available in the presence of new sizable contributions to the EWPT. However we will not consider this possibility
in the following.} Moreover, a promising excess has been found by both experiments at $m_H\simeq 125$~GeV,
a reasonable expectation is that this excess will be confirmed by the 2012 LHC data and the Higgs will be
finally discovered.

Very little can be said, on the contrary, about the \emph{nature} of the Higgs. Minimality suggests that it should
be an elementary weakly coupled particle, described by the Higgs model up to very high energy scales.
Naturalness requires instead that a more complicated structure emerges around the TeV.
A natural Higgs could still be elementary, if it is embedded in a supersymmetric framework, or it could be a
\emph{composite} object, {\it{i.e.}} the bound state of a new strong dynamics. In the latter case
the Hierarchy Problem is solved by the finite size of the Higgs, which screens the contributions
to its mass from virtual quanta of
short wavelength. A particularly plausible possibility is that the Higgs is not a generic
bound state of the strong dynamics, but rather
a (pseudo) Nambu-Goldstone Boson (pNGB) associated to a spontaneously broken symmetry.
This explains naturally why it is much lighter than the other, unobserved, strong sector resonances.

The idea of a composite pNGB Higgs has been studied at length \cite{gk,Agashe:2004rs,Contino:2006qr,Contino:2006nn,Giudice:2007fh,Panico:2011pw,BS,CHM2,CHM2flat,Panico:2006em},
and the following scenario has emerged. Apart from the Higgs, which is composite, all the other
SM particles originate as \emph{elementary}
fields, external to the strong sector. The communication with the Higgs and the strong dynamics,
and thus the generation of masses after ElectroWeak Symmetry Breaking (EWSB),
occurs through \emph{linear} couplings of the elementary fields with suitable strong sector operators.
At low energies, below the confinement scale, the linear couplings
become \emph{mixing terms} among the elementary SM particles and some heavy composite resonance.
The physical states after diagonalization possess a composite
component, realizing the paradigm of ``partial compositeness'' \cite{Kaplan:1991dc, Contino:2006nn}. The simplest setup,
which is almost universally adopted in the literature,
is based on the ${\textrm{SO}}(5)\rightarrow{\textrm{SO}}(4)$ symmetry breaking pattern.
This delivers only one pNGB Higgs doublet and incorporates the custodial
 symmetry to protect $\delta\rho$ from unacceptably large corrections. A possible model-building
ambiguity comes from the choice of the ${\textrm{SO}}(5)$
 representation of the fermionic operators that mix with the SM fermions and in particular
with the third family $q_L$ and $t_R$. In the ``minimal'' scenario, often denoted
 as ``MCHM5'' in the literature, these operators are in the fundamental (${\mathbf{5}}$)
representation. Other known possibilities are the ${\mathbf{4}}$ (MCHM4)
 and the ${\mathbf{10}}$ representations. Even if in the present paper we will adopt the
minimal scenario as a reference, our results have a more general
 valence and range of applicability. We will discuss in the Conclusions the effect of changing
the representations and of even more radical deformations.

 Associated with the fermionic operators, massive colored fermionic resonances emerge
from the strong sector. These are the so-called ``top partners'' and
 provide a very promising direct experimental manifestation of the composite Higgs
scenario at the LHC \cite{topp}. Indeed it has been noticed by many authors,
 and in Ref.~\cite{Contino:2006qr} for the first time, that in explicit concrete models
these particles are anomalously light, much lighter than the other strong
 sector's resonances. Concretely, one finds that the partners can easily be below $1$~TeV,
with an upper bound of around $1.5$~TeV, while the
 typical strong sector's scale is above $3$~TeV in order to satisfy the EWPT constraints.
Moreover, Ref.~\cite{Contino:2006qr} observed a certain correlation
 of the mass of the partners with the one of the Higgs boson.

 The first goal of our paper will be to show that the lightness of the top partners
has a \emph{structural} origin, rather than being a peculiarity of some explicit
 model. The point is that in the composite Higgs scenario there is a tight relation
 among the top partners and the generation of the Higgs potential. This leads
 to a \emph{parametric} correlation among the mass of the partners and the one of the
Higgs boson. In order for the latter to be light as implied by the present
 data, we find that at least one of the top partners must be anomalously light.
In section~\ref{sec:higgspartners} we will describe this mechanism in detail by adopting
a general description of the composite Higgs scenario with partial compositeness developed
in ref.s~\cite{Agashe:2004rs,Contino:2006nn,Giudice:2007fh}. Our results will thus have general
validity and they will apply, in particular, to the 5d holographic models of ref.s~\cite{Agashe:2004rs,Contino:2006qr}.

For a quantitative confirmation of the effect we need to
study a concrete realization
of the composite Higgs idea. The simplest possibility is to consider
 a ``Discrete'' Composite Higgs Model (DCHM) like the one proposed by two of us in
Ref.~\cite{Panico:2011pw}. The central observation behind the formulation
 of the DCHM is that the potential of the composite Higgs is saturated by the IR
dynamics of the strong sector. Indeed in the UV, above the mass
 of the resonances which corresponds to the confinement scale, the Higgs ``dissolves''
in its fundamental constituents and the contributions to the potential
 get screened, as mentioned above. The same screening must take place in the low energy
effective description of the theory and therefore the dynamics of the
 resonances must be such to give a \emph{finite} and \emph{calculable} ({\it{i.e.}},
IR-saturated) Higgs potential. This is indeed what happens in the 5d
 holographic models thanks to the collective effect of the entire Kaluza-Klein tower.
In the DCHM instead this is achieved by introducing a finite number of
 resonances and of extra symmetries which realize a ``collective breaking'' \cite{ArkaniHamed:2001ca}
protection of the Higgs potential. Further elaborations
 based on this philosophy can be found in Ref.~\cite{DeCurtis:2011yx}. Other models similar
to the DCHM have been proposed in the context of Little-Higgs
 theories \cite{Cheng:2006ht}.

 In sections~3 and 4 we describe in detail the structure of the top partners in the DCHM,
and derive analytic explicit formulas that show quantitatively the
 correlation with the Higgs mass. Section~3 is devoted to the study of the 3-site DCHM,
which provides a genuinely complete theory of composite Higgs. In this
 model, two layers of fermionic resonances are introduced and the Higgs potential is
completely finite. In section~4 we consider instead a simpler but less complete
 model, the two-site DCHM. In this case one has a single layer of resonances and quite
a small number ($3$, after fixing the top mass) of parameter describing
 the top partners. However the potential is \emph{not} completely calculable, being
affected by a logarithmic divergence at one loop \cite{Panico:2011pw}.
 Nevertheless it turns out that the divergence corresponds to a \emph{unique} operator
in the potential and therefore it can be canceled by renormalizing
 only one parameter which we can chose to be the Higgs VEV $v$. Thus, the Higgs
mass \emph{is} calculable also in the DCHM$_2$, this model can therefore
 be considered as the ``simplest'' composite Higgs model and it can be used to study
the phenomenology of the top partners in correlation with the Higgs mass.

 The analytic results are further supported by scatter plots, in which we scan all
the available parameter space of the model. The results are quite remarkable:
 in the plane of the masses of the top partners the points with light enough Higgs
boson fall very sharply in the region of light partners. Notice that the actual
 values of the partner's mass is \emph{not} fixed by our argument, it still depends
on the overall mass scale of the strong sector. However this scale can only be
 raised at the price of fine-tuning the parameter $\xi\simeq (v/f_\pi)^2$ (as defined
in Ref.~\cite{Giudice:2007fh}) to very small values. Reasonable values of $\xi$,
 below which the entire scenario starts becoming  unplausible, are $\xi=0.2$ or
$\xi=0.1$. For $\xi=0.2$ we find that the partners are \emph{always} below
 $1$~TeV while for $\xi=0.1$ the absolute maximum is around $1.5$~TeV. We therefore
expect that the $14$-TeV LHC will have enough sensitivity to explore
 the parameter space of the model completely.

 Non-trivial constraints can however already be obtained by the presently available
exclusions from the $7$-TeV data, as we will discuss in section~5. The main
 effect comes from the exclusion of the 5/3-charged partner decaying to $tW^+$.
No dedicated LHC search is available for this particle, however we find that it is
  possible to apply the bound of $611$~GeV coming from the CMS search of bottom-like
heavy quarks in same-sign dileptons \cite{CMS:BtoWt}. The bounds on the other partners
are considerably reduced by the branching fractions to the individual decay channels assumed
in the searches. We therefore expect that a significant improvement of the bounds could be obtained
within some explicit model by combining the different channels.
The $2$-site DCHM is definitely the best candidate, one might easily perform a complete scan of its
$3$ free parameters. We have already implemented  the $2$-site DCHM (and also of the $3$-site one)
in {\textsc{MadGraph}} 5 \cite{Alwall:2011uj} for the required simulations.

Finally, in section~6 we present our conclusion and an outlook of the possible implication
of our results. In particular we discuss how our analysis could be adapted
to more general scenarios of composite Higgs and we suggest some directions
for future work.

\section{Light Higgs wants light partners}\label{sec:higgspartners}

If the Higgs is a pNGB its potential, and in particular its mass $m_H$, can only
be generated through the breaking of the Goldstone symmetry.
One unavoidable, sizable source of Goldstone symmetry breaking is the top quark Yukawa
coupling $y_t$. Thus it is very reasonable to expect a tight relation
among the Higgs mass and the fermionic sector of the theory which is responsible
for the generation of $y_t$. This is particularly true in the canonical scenario
of composite Higgs, summarized in the Introduction, where the \emph{only} sizable contribution to the Higgs potential
comes from the top sector. In more general cases there might be
additional terms, coming for instance from extra sources of symmetry breaking not
associated with the SM fermions and gauge fields \cite{luty}. Barring fine-tuning,
the latter contributions can however at most \emph{enhance} $m_H$, the ones from
the top therefore provide a robust \emph{lower bound} on the Higgs mass.
If the Higgs has to be light, as it seems to be preferred by the present data,
the top sector contribution must therefore be kept small enough by some mechanism.
In the minimal scenario, as we will describe below, this is achieved by making anomalously
light (and thus more easily detectable) some of the exotic states in the top sector.

In order to understand this mechanism we obviously need to specify in some detail
the structure of the theory which controls the generation
of $m_H$ and $y_t$. As anticipated in the Introduction,
the paradigm adopted in the minimal model is the one of
partial compositeness, in which the elementary
left- and right-handed top fields are mixed with heavy vector-like colored particles,
the so-called top partners. After diagonalization the physical top
becomes an admixture of elementary and composite states and interacts with the strong sector,
and in particular with the Higgs, through its composite
component. The Yukawa coupling gets therefore generated and it is proportional
to the sine of the mixing angles $\varphi_{L,R}$. The
relevant Lagrangian, introduced in ref.~\cite{Contino:2006nn},
has the structure
\begin{eqnarray}
&&{\mathcal{L}}_{\textrm{mass}}\,=\,-\left(y_L f_\pi \,\overline{t}_LT_R+y_R f_\pi\, \overline{t}_R {\widetilde{T}}_L+{\textrm{h.c.}}\right)-m_T^*\overline{T}T-m_{\widetilde{T}}^*\overline{{\widetilde{T}}}{\widetilde{T}}\,,\nonumber\\
&&{\mathcal{L}}_{\textrm{Yuk}}\,=\,Y_*h\overline{T}  {\widetilde{T}}+{\textrm{h.c.}}\,,
\label{eqm1}
\end{eqnarray}
where $h$ is the Higgs field (before EWSB, {\it{i.e.}} $h=v+\rho$) and we have employed the decay constant
$f_\pi$ of the Goldstone boson Higgs for the normalization
of the elementary--composite mixings. After diagonalization, neglecting EWSB, the top Yukawa reads
\begin{equation}
y_t\,=\,Y_*\sin\varphi_{L}\sin\varphi_{R}\,,\;\;\;\;\;{\textrm{with}}\;\;\;\;\;\;\;\;\left\{
\begin{array}{l}\sin\varphi_{L}=\frac{y_L f_\pi}{m_T}\\
\sin\varphi_{R}=\frac{y_R f_\pi}{m_{\widetilde{T}}}
\end{array}\right.\,,
\label{eq0}
\end{equation}
where $m_{T,{\widetilde{T}}}=\sqrt{(m_{T,{\widetilde{T}}}^*)^2+(y_{L,R}f_\pi)^2}$
are the physical masses of the top partners.~\footnote{Actually, the physical masses receive extra tiny corrections due to EWSB.}

The essential point of making the partners light is that this allows to
\emph{decrease} the elementary--composite mixings $y_{L,R}$ while
keeping $y_t$ fixed to the experimental value. Let us consider the case
of comparable left- and right-handed mixings, $y_L\simeq y_R\equiv y$.
This condition, as explained in the following (see also  \cite{Panico:2011pw}),
is enforced in the minimal model by the requirement of a realistic EWSB.
We can also assume that $m_{T}^*$ and $m_{{\widetilde{T}}}^*$, while potentially
small, are still larger than $y_{L}f_\pi$ and $y_{R}f_\pi$, the critical value
after which eq.~(\ref{eq0}) saturates and there is no advantage in further
decreasing the masses. Under these conditions eq.~(\ref{eq0}) gives
\begin{equation}
\displaystyle
y^2\,=\,\frac{y_t}{Y_*}\frac{m_Tm_{\widetilde{T}}}{f_\pi^2}\,,
\label{eq1}
\end{equation}
which shows how $y^2$ decreases \emph{linearly} with the mass of each partner.

The mixings ensure the communication among the strong sector, which is invariant
under the Goldstone symmetry, and the elementary
sector which is not. Therefore they break the symmetry and allow for the generation
of the Higgs mass. It is thus intuitive that a reduction of their value,
as implied by eq.~(\ref{eq1}) for light top partners, will lead to a decrease of $m_H$.
To be quantitative, let us anticipate the result of the following section
(see also \cite{Contino:2006qr} and \cite{Panico:2011pw}): $m_H$ can be estimated as
\begin{equation}
m_H\simeq \sqrt{\frac{3}{2}}\frac{y^2 v}{\pi}\,,
\label{eq2}
\end{equation}
where $v\simeq246$ GeV is the Higgs VEV. This gives, making use of eq.~(\ref{eq1})
\begin{equation}
m_H\simeq 4 \sqrt{3}m_t\,\frac{m_Tm_{\widetilde{T}}}{4\pi Y_* f_\pi^2}\,.
\label{eq3}
\end{equation}
The above equation already shows the correlation among the Higgs and the top partner
mass. Of course we still need to justify eq.~(\ref{eq2}) and
for this we need the more detailed analysis of the following section.

There is however one important aspect which is \emph{not} captured by this general
discussion. We see from eqs.~(\ref{eq1}) and (\ref{eq3}) that
making both $m_T$ and $m_{\widetilde{T}}$ small at the same time produces a
\emph{quadratic} decrease of $y^2$ and thus of $m_H$. However this behavior
is never found in the explicit models we will investigate in the following sections,
the effect is always \emph{linear}. The basic reason is that, due to the
Goldstone nature of the Higgs, the coupling $Y_*$ defined in eq.~(\ref{eqm1}) depends
itself on the partners mass. Indeed all the interactions of a pNGB Higgs are controlled by the
dimensional coupling $f_\pi$ and no independent Yukawa-like coupling $Y_*$  can emerge.
By dimensional analysis on has $Y_*\simeq m_{T,\widetilde{T}}^*/f_\pi$ or more precisely,
as we will also verify below,
$Y_*\simeq{\textrm{max}}(m_{T}^*,m_{\widetilde{T}}^*)/f_\pi$. Thus if both masses become
small one power of $m_{T,{\widetilde{T}}}$ in eqs.~(\ref{eq1}) and (\ref{eq3}) is
compensated by $Y_*$ and the effect remains linear.

\subsection{General analysis}\label{sec:genan}

For a better understanding we need a slightly more careful description of our theory.
In particular we must take into account the Goldstone boson
nature of the Higgs
which is instead hidden in the approach of ref.~\cite{Contino:2006nn} adopted in the
previous discussion.
Following ref.s~\cite{Agashe:2004rs,Giudice:2007fh} (see also \cite{BS}) we describe the Higgs as a pNGB
associated with the ${\textrm{SO}}(5)\rightarrow {\textrm{SO}}(4)$ spontaneous
breaking which takes place in the strong sector.
We parametrize (see \cite{Panico:2011pw} for the conventions) the Goldstone boson matrix as
\begin{equation}
\displaystyle
U=e^{i\frac{\sqrt{2}}{f_\pi}\Pi_{\widehat{a}}T^{\widehat{a}}}\,,
\label{go}
\end{equation}
where $T^{\widehat{a}}$ are the broken generators and $\Pi_{\widehat{a}}$ the $4$
real Higgs components. The Goldstone matrix transforms
under $g\in {\textrm{SO}}(5)$ as \cite{Coleman:1969sm}
\begin{equation}
U\;\rightarrow\;g\cdot U\cdot h^t\left(\Pi;\,g\right)\,,
\label{Utrasf}
\end{equation}
where $h$ is a non-linear representation of ${\textrm{SO}}(5)$ which however
only lives in ${\textrm{SO}}(4)$. With our choice of the generators
$h$ is block-diagonal
\begin{equation}
h\,=\,\left(\begin{array}{cc}
h_4\,&0\\
0&1
\end{array}\right)\,,
\label{h}
\end{equation}
with $h_4\in {\textrm {SO}}(4)$

The SM fermions, and in particular the third family quarks $q_L=(t_L\;b_L)$ and
$t_R$, are introduced as elementary fields and they are coupled
linearly to the strong sector. In the UV, where ${\textrm{SO}}(5)$ is restored,
we can imagine that the elementary--composite interactions take the form
\begin{eqnarray}
{\mathcal{L}}\,&&=\,y_L\left({\overline{q}}_L\right)^\alpha\Delta_{\alpha I}^L\left({\mathcal{O}}_R\right)^I\,+
\,y_R\left({\overline{t}}_R\right)\Delta_{I}^R\left({\mathcal{O}}_L\right)^I\,+\,\textrm{h.c.}\,,
\label{mix}
\end{eqnarray}
where the chiral fermionic operators ${\mathcal{O}}_{L,R}$ transform in a
\emph{linear} representation of ${\textrm{SO}}(5)$.  \footnote{We have defined
the mixings $y_{L,R}$ as dimensionless couplings, for shortness we have reabsorbed
in ${\mathcal{O}}_{L,R}$  the powers of the UV scale needed to
restore the correct energy dimensions.}
In particular in the minimal
model we take both ${\mathcal{O}}_{L}$ and ${\mathcal{O}}_{R}$ in the fundamental,
${\mathbf5}$. The tensors $\Delta^{L,R}$ are uniquely
fixed by the need of respecting the SM ${\textrm{SU}}(2)\times{\textrm{U}}(1)_Y$
group embedded in ${\textrm{SO}}(5)$ \footnote{Actually,
one extra ${\textrm{U}}(1)_X$ global factor is needed. In order to reproduce the
correct SM hypercharges one must indeed define $Y=X+T^3_R$
and assign $2/3$ ${\textrm{U}}(1)_X$ charge to both ${\mathcal{O}}_{L}$ and ${\mathcal{O}}_{R}$.\label{foot}}
\begin{eqnarray}
&&\Delta^L_{\alpha I}\,=\,\frac1{\sqrt{2}}\left(\begin{array}{ccccc}\;0&\;0&\;1&-i&\;0\\\;1&\;i&\;0&\;0&\;0\end{array}\right)\,,\nonumber\\
&&\Delta^R_I\,=\,-i\left(0\;0\;0\;0\;1\right)\,.
\label{spu}
\end{eqnarray}
Let us also define, for future use, the embedding in the ${\mathbf5}$ of $q_L$ and of $t_R$
\begin{eqnarray}
&&\left(q_L^{\mathbf5}\right)^I\,=\,\left({\Delta^L}^*\right)^{\alpha I}\left(q_L\right)_\alpha=
\frac1{\sqrt{2}}\left(\begin{array}{ccccc}b_L&-i b_L&\;\;t_L&\;i t_L&\;\;0\end{array}\right)\,,\nonumber\\
&&\left(t_R^{\mathbf5}\right)^I\,=\,\left({\Delta^R}^*\right)^{ I} t_R=i
\left(\begin{array}{ccccc}0&0&0&0&t_R\end{array}\right)\,.
\label{emb}
\end{eqnarray}

The elementary--composite couplings obviously break the Goldstone symmetry
${\textrm{SO}}(5)$. However provided the breaking is small we can still obtain
valuable information from the  ${\textrm{SO}}(5)$ invariance by the method of spurions.
The point is that the theory, including the UV mixings in
eq.~(\ref{mix}), is perfectly invariant if we transform not just the strong sector
fields and operators but also the tensors $\Delta^L$ and
$\Delta^R$. This invariance survives in the IR description, the effective operators
must therefore respect ${\textrm{SO}}(5)$ if we treat $\Delta^L$ and $\Delta^R$ as spurions which transform,
formally, in the ${\mathbf5}$ of ${\textrm{SO}}(5)$. To be precise there are further
symmetries one should take into account. These are the ``elementary''
${\textrm{U}}(2)^0_L$ and ${\textrm{U}}(1)^0_R$, under which the strong sector is
invariant and only the elementary fermions and the spurions transform.
Certain linear combinations of the elementary group generators with the
${\textrm{SO}}(5)$ (and ${\textrm{U}}(1)_X$, see footnote~\ref{foot}) ones
correspond to the SM group, these are of course preserved by the mixings.

\subsubsection*{The Higgs potential}

Let us first discuss the implications of the spurionic analysis on the structure of
the Higgs potential. We must classify the non-derivative invariant operators
involving the Higgs and the spurions. Notice that the invariance under
${\textrm{U}}(2)^0_L\times{\textrm{U}}(1)^0_R$ requires that the spurions only appear
in the following two combinations
\begin{eqnarray}
&&\Gamma^L_{IJ}\,=\,\left({\Delta^L}^*\right)^\alpha_I\left(\Delta^L\right)_{\alpha J} \,,\nonumber\\
&&\Gamma^R_{IJ}\,=\,\left({\Delta^R}^*\right)_I\left(\Delta^R\right)_{J} \,.
\end{eqnarray}
The Higgs enters instead through the Goldstone matrix $U$. Notice that to
build ${\textrm{SO}}(5)$ invariants we must contract the indices of $\Gamma^{L,R}$ with
the \emph{first} index of the matrix $U$, and not with the second one.
Indeed if we rewrite more explicitly equation (\ref{Utrasf}) as
\begin{equation}
U_{I\bar{J}}\;\rightarrow\;g_{I}^{\;\;I'}U_{I'{\bar{J}}'}h_{\bar{J}}^{\;\;\bar{J}'}\,,
\end{equation}
we see that while the first index transforms with $g$ like the spurion indices do,
the second one transforms differently, with $h$. Remember that $h$ is block-diagonal
(see eq.~(\ref{h})), thus to respect the symmetry we just need to form ${\textrm{SO}}(4)$
(rather than ${\textrm{SO}}(5)$) invariants with the ``barred'' indices,
in practice we can split them in fourplet and singlet components as $\bar{I}=\{i,\,5\}$.

With these tools it is straightforward to classify all the possible invariants
at a given order in the spurions. At the quadratic order, up to irrelevant additive constants,
only two independent operators exist
\begin{eqnarray}
&&v^L(h)\,=\,\left(U^t\cdot\Gamma^L\cdot U\right)_{55}\,=\,\frac12\sin^2{h/f_\pi}\,,\nonumber\\
&&v^R(h)\,=\,\left(U^t\cdot\Gamma^R\cdot U\right)_{55}\,=\,\cos^2{h/f_\pi}\,=\,1-\sin^2{h/f_\pi}\,,
\label{eq:oper}
\end{eqnarray}
where we plugged in the explicit value of the spurions in eq.~(\ref{spu}) and
of the Goldstone matrix in eq.~(\ref{go}) taking the Higgs along its VEV
$\langle\Pi^{\widehat{a}}\rangle=h\delta^{\widehat{a}4}$. At this order then
the potential can only be formed by two operators, with unknown coefficients which
would become calculable only within an explicit model.
We can nevertheless estimate their expected size. Following \cite{Giudice:2007fh,Mrazek:2011iu} we obtain
\begin{equation}
V^{(2)}(h)\,=\,\frac{N_cM_*^4}{16\pi^2g_*^2}\left[c_Ly_L^2v^L(h)+c_Ry_R^2v^R(h)\right]\,=\,\frac{N_cM_*^4}{16\pi^2g_*^2}\left[\frac12 c_Ly_L^2-c_Ry_R^2\right]\sin^2(h/f_\pi)\,+\,{\textrm{const.}}\,,
\label{pot2}
\end{equation}
where $c_{L,R}$ are order one parameters and $\{M_*,\,g_*\}$ are the typical
masses and couplings of the strong sector, $g_*$ is defined as
$g_*= M_*/f_\pi$. Remember that what we are discussing is the fermionic
contribution to the potential, generated by colored fermion loops, this
is the origin of the $N_c=3$ QCD color factor in eq.~(\ref{pot2}). Also,
this implies that the scale $M_*$ is the one of the fermionic resonances,
which could be a priori different from the mass of the vectors
$m_\rho$. \footnote{The mass $M_*$ is the scale at which the potential is saturated
and generically it \emph{is not} associated to the masses $m_{T,{\widetilde{T}}}$ of the
anomalously light partner. Due to additional structures, and only in the case in which \emph{both}
 $T$ and  ${{\widetilde{T}}}$ are anomalously light, one might obtain $M_*\sim m_{T,{\widetilde{T}}}$
 in some explicit model because the light degrees of freedom reconstruct the structure of a $2$-site DCHM
 in which the quadratic divergence is canceled.
}

The spurionic analysis has strongly constrained the Higgs potential at the
quadratic order. The two independent operators have indeed the same functional dependence on the Higgs
and the potential is entirely proportional to $\sin^2(h/f_\pi)$.
But then the potential at this order cannot lead to a
realistic EWSB, the minimum is either at $h=0$ or at $h=\pi f_\pi/2$. We would instead need
to adjust the minimum in order to have $\xi=\sin^2(v/f_\pi)<1$, and to achieve
this additional contributions are required. In the minimal
scenario these are provided by higher order terms in the spurion expansion.
The classification of the operators is straightforwardly extended to the
quartic order, one finds a second
allowed functional dependence \footnote{Actually, also a term proportional
to $\cos{h/f_\pi}$ could appear. This is however forbidden by the parity in
$SO(4)$, $P_{LR}$, for this reason it is not present in the minimal models.}
\begin{equation}
V^{(4)}(h)\,=\,\frac{N_cM_*^4}{16\pi^2 g_*^4}\left[c^{(4)}_1\,y^4\sin^2(h/f_\pi)+c^{(4)}_2\,y^4\sin^2(h/f_\pi)\cos^2(h/f_\pi)\right]\,,
\label{potfin}
\end{equation}
where $y^4$ collectively  denotes the quartic terms $y_L^4$, $y_R^4$ or
$y_L^2y_R^2$ and $c_{1,2}^{(4)}$ are coefficients of order unity.
Notice that, differently from the quadratic one, the quartic potential
does not depend strongly on the fermionic scale $M_*$. Since $M_*= g_*f_\pi$
the prefactor of $V^{(4)}$ can indeed be rewritten as $f_\pi^4$.

A priori, $V^{(4)}$ should give a negligible contribution to the potential because it
is suppressed with respect to $V^{(2)}$ by a factor $(y_{L,R}/g_*)^2$, which is small
in the minimal scenario. To achieve realistic EWSB however we need to tune
the coefficients of the $\sin^2(h/f_\pi)$ and $\sin^2(h/f_\pi)\cos^2(h/f_\pi)$
terms in such a way as to cancel the Higgs mass term obtaining $v/f_\pi<1$.
In formulas, we have
\begin{equation}
V\,=\,\alpha\sin^2(h/f_\pi)\,-\,\beta\sin^2(h/f_\pi)\cos^2(h/f_\pi)\,,\;\;\;\;\;
\Rightarrow\;\;\;\;\;\sin^2(v/f_\pi)=\frac{\beta-\alpha}{2\beta}\ll1\,.
\label{tun0}
\end{equation}
But, to make $\alpha\simeq\beta$, we need to cancel $V^{(2)}$, which only
contributes to $\alpha$ and not to $\beta$, and to make it comparable with $V^{(4)}$.
This requires $y_L\simeq y_R\equiv y$ or, more precisely
\begin{equation}
\frac12c_Ly_L^2=c_Ry_R^2\left(1+{\mathcal{O}}(y^2/g_*^2)\right)\,.
\label{rel}
\end{equation}
On top of this preliminary cancellation the tuning of the Higgs VEV in
eq.~(\ref{tun0}) must be carried on. The total amount of fine-tuning is of order
\begin{equation}\label{tunn}
\left(\frac{y}{g_*}\right)^2\,\sin^2(v/f_\pi) =
\left(\frac{y}{g_*}\right)^2\,\xi\,,
\end{equation}
and it is worse than the naive estimate by the factor
$(y/g_*)^2$. \footnote{The theory would then be more natural if
$y\sim g_*$. For small values of $g_*$, however, all the
fermionic resonaces become lighter and this could give rise to
enhanced corrections to the electroweak parameters in contrast with the EWPT.
It could however be interesting to study this case
explicitly in a concrete model.}
\footnote{We remind the reader that the results of the presence section have general validity, in particular they apply to the 5d
holographic models studied at length in the literature. In that context the need of an enhanced tuning in order to obtain
realistic EWSB has been already pointed out \cite{Panico:2006em} by explicitly computing the logarithmic derivative.}

The final outcome of this discussion is that achieving realistic EWSB
requires that the quadratic potential is artificially reduced and made comparable with
$V^{(4)}$. Therefore we can simply forget about $V^{(2)}$ in eq.~(\ref{pot2})
and use instead eq.~(\ref{potfin}) as an estimate of the total Higgs potential.
In particular we can estimate the physical Higgs mass, which is given by
\begin{equation}
m_H^2\,=\,\frac{8\beta}{f_\pi^2}\sin^2(v/f_\pi)\cos^2(v/f_\pi)\,\simeq\,\frac{2 N_c y^4}{16\pi^2}f_\pi^2\sin^2(2v/f_\pi)\,,
\label{mh}
\end{equation}
where we used $g_*= M_*/f_\pi$. Expanding for $v/f_\pi\ll1$ we recover the result anticipated in eq.~(\ref{eq2}).

We would have reached very similar conclusions if we had considered fermionic operators in the ${\mathbf4}$ of
${\textrm{SO}}(5)$ rather than in the ${\mathbf5}$. As shown in the original paper on the minimal composite Higgs
\cite{Agashe:2004rs}, also in that case the potential is the sum of two trigonometric functions with coefficients $\alpha$
and $\beta$ that scale respectively as $\alpha\sim y^2$ and $\beta\sim y^4$. The condition to obtain a realistic EWSB
is again $\alpha\simeq\beta$, {\it{i.e.}} eq.~(\ref{rel}), therefore the Higgs mass-term scales like $y^4$ as in eq.~(\ref{mh}).
Since the scaling is the same, all the conclusions drawn in this section, in particular the main result in eq.~(\ref{eq:mhgen}), will
hold in exactly the same way. Moreover it is possible to show that the same structure of the potential emerges in the
case of the ${\mathbf{10}}$ of ${\textrm{SO}}(5)$ so that our results will apply also to the latter case. Possible ways to
evade the light Higgs-light partner correlation of eq.~(\ref{eq:mhgen}) will be discussed in the Conclusions.

\subsubsection*{The top mass}

For a quantitative estimate of $m_H$, which will show the correlation
with the top partners mass, we need an estimate of $y$. The mixings
$y_{L,R} $ control the generation of the top quark Yukawa, which of
course must be fixed to the experimental value. The size of $y$ however is
not uniquely fixed because $y_t$ also depends on the masses of the top
partners with which the elementary $t_L$ and $t_R$ fields mix. In particular,
as explained previously (see eq.~(\ref{eq0})), the top Yukawa would
get \emph{enhanced} in the presence of anomalously light partners. To compensate
for this, while keeping $y_t$ fixed, one has to \emph{decrease} $y$, thus lowering the Higgs mass.

We can study this effect in detail by writing down the low energy effective
Lagrangian for the top partners. Since the operators
${\mathcal{O}}_{L,R}$ are in the ${\mathbf{5}}$ of ${\textrm{SO}}(5)$,
which decomposes as ${\mathbf{5}}={\mathbf{4}}\oplus{\mathbf{1}}$
under ${\textrm{SO}}(4)$, the top partners which appear in the low energy
theory will be in the fourplet and in the singlet. \footnote{Of course many more
states could exist, associated to other UV operators. The presence of the
fourplet and the singlet seems however unavoidable.} We describe these states
as CCWZ fields, which transform non-linearly under
${\textrm{SO}}(5)$ \cite{Coleman:1969sm}. In particular the fourplet transforms as
\begin{equation}
Q_i\;\rightarrow\;\left(h_4\right)_i^{\;\;j}Q_j\,,
\end{equation}
with $i=1,\ldots4$ and $h_4$ as in eq.~(\ref{h}). The singlet ${\widetilde{T}}$
is obviously invariant.
For our discussion we will not need to write down the complete Lagrangian,
but only the mass terms and mixings. We classify the operators with the spurion
method previously outlined and we find, at the leading order
\begin{eqnarray}
{\mathcal{L}}\,=\,&&-m_T^*\overline{Q} Q-m_{\widetilde{T}}^*\overline{\widetilde{T}} \widetilde{T} \nonumber\\
&&- y_L f_\pi \left({\bar{q}_L^{(5)}}\right)^I\left(a_LU_{Ii}Q_R^i+b_LU_{I5}{\widetilde{T}}_R\right)+{\textrm{h.c.}}\nonumber\\
&&- y_R f_\pi \left({\bar{t}_R^{(5)}}\right)^I\left(a_RU_{Ii}Q_L^i+b_RU_{I5}{\widetilde{T}}_L\right)+{\textrm{h.c.}}\,,
\label{lagt}
\end{eqnarray}
where the embeddings $q_L^{\mathbf5}$ and $t_R^{\mathbf5}$ are defined in eq.~(\ref{emb}).

The one in eq.~(\ref{lagt}) is the most general fermion mass Lagrangian
allowed by the ${\textrm{SO}}(5)$ Goldstone symmetry, it is not
difficult to see that it leads to a top mass
\begin{equation}
m_t \simeq \frac{|b_L^* b_R m_T^*-a_L^* a_R m_{\widetilde{T}}^*|}
{2 \sqrt{2} |a_L||b_R|}\sin \varphi_L \sin \varphi_R
\sin\left(2 v/{f_\pi}\right)\,,\;\;\;\;\;{\textrm{with}}\;\;\;\;\;\;\;\;\left\{
\begin{array}{l}\sin\varphi_{L}=\frac{|a_L| y_L f_\pi}{m_T}\\
\sin\varphi_{R}=\frac{|b_R| y_R f_\pi}{m_{\widetilde{T}}}
\end{array}\right.\,,
\label{mtgen}
\end{equation}
where $m_{T}^2={(m_{T}^*)^2+|a_L|^2 y_L^2 f_\pi^2}$ and
$m_{\widetilde{T}}^2={(m_{\widetilde{T}}^*)^2+|b_R|^2 y_R^2 f_\pi^2}$ are the
physical masses of the partners before EWSB. Making contact with eq.~(\ref{eq0})
we find, as anticipated, that the Yukawa is controlled by the masses:
$Y_*\simeq |b_L^* b_R m_T^*-a_L^* a_R m_{\widetilde{T}}^*|/f_\pi$.

Barring fine-tuning and assuming  $m^*_{T,{\widetilde{T}}}\simeq m_{T,{\widetilde{T}}}$ we can approximate
\begin{equation}
m_t \simeq \frac{{\textrm{max}}(m_T^*, m_{\widetilde{T}}^*)}
{2 \sqrt{2} }\sin \varphi_L \sin \varphi_R
\sin\left(2 v/{f_\pi}\right)
\,=\,
 \frac{1}{2 \sqrt{2}}\frac{y_L y_R f_\pi^2}{{\textrm{min}}(m_T, m_{\widetilde{T}})}
\sin\left(2 v/{f_\pi}\right)\,.
\label{yyt}
\end{equation}

\subsubsection*{Light partners for a light Higgs}

The equation above, combined with the formula (\ref{mh}) for $m_H$ finally shows the correlation among the
Higgs and the top partners mass:
\begin{equation}
m_H\,\simeq\,\frac{\sqrt{N_c}}{\pi}\frac{{\textrm{min}}(m_T, m_{\widetilde{T}})}{f_\pi}m_t\,\simeq\,130\,{\textrm{GeV}}\frac{{\textrm{min}}(m_T, m_{\widetilde{T}})}{1.4f_\pi}\,.
\label{eq:mhgen}
\end{equation}
For $f_\pi\simeq500$~GeV we see that satisfying the LHC bound on
$m_H$ of around $130$~GeV requires the presence of at least one state of mass below $700$~GeV.
\footnote{Given that the Higgs is composite it has modified coupling and therefore we cannot apply directly the SM exclusions. The upper bound of
$130$~GeV takes into account the effects of compositeness as we will discuss in Section~3.3.}
For
$f_\pi\simeq750$~GeV, which already corresponds to a significant
level of fine-tuning, the partners can reach $1$~TeV. This estimate suggests that
the requirement of a realistic Higgs mass forces the theory to deliver
relatively light top partners, definitely within the
reach of the $14$~TeV LHC and possibly close to the present bounds from
the run at $7$~TeV. We will support this claim in the following sections
where we will analyze the top partners spectrum within two explicit models.

The existence of an approximate linear correlation among $m_H$ and the mass
of the lightest top partner $m_{\textrm{light}}={\textrm{min}}(m_T, m_{\widetilde{T}})$
was already noticed in ref.~\cite{Contino:2006qr} in the case of holographic
models, however the physical interpretation of the result was not properly
understood. To make
contact with the argument presented in \cite{Contino:2006qr}, we notice that from
a low-energy perspective the Higgs mass-term arises from a quadratically divergent loop
of elementary fermion fields, mixed with strength $y=y_{L,R}$ to the strong sector as in
eq.~(\ref{mix}). One can estimate
$$
m_H^2\sim \frac{N_c}{16\pi^2}\xi\frac{y^4}{g_*^2}\Lambda^2\,,
$$
where $\Lambda$ denotes the cutoff scale of the loop integral. To account for the
observed linear relation among $m_H$ and $m_{\textrm{light}}$, ref.~\cite{Contino:2006qr}
claims that $\Lambda\simeq m_{\textrm{light}}$, {\it{i.e.}} that the propagation
of the lightest top partner in the loop is already sufficient to cancel the quadratic
divergence.
If the pre-factor $(y^2/g_*)^2$ can be estimated with the naive
partial-compositeness relation $y_t=y^2/g_*$, irregardless of the presence of the
light partner,  by assuming $\Lambda\simeq m_{\textrm{light}}$ one recovers
eq.~(\ref{eq:mhgen}).
However this argument is incorrect for two reasons. First of all the presence of anomalously
light partners modifies the naive relation among $y$ and $y_t$. This is obvious because
the elementary-composite mixing angle, and thus $y_t$, must be
enhanced if the composite particle becomes
light. One finds indeed $y_t=y^2 f/m_{\textrm{light}}$ as shown in eq.~(\ref{yyt}). Moreover there
is no reason why the cutoff scale $\Lambda$ should be set by $m_{\textrm{light}}$. Indeed
there is no known mechanism through which a single multiplet of ${\textrm{SO}}(4)$
(the four-plet $Q$ or the singlet ${\widetilde{T}}$) could cancel the quadratic divergence
of $m_H$, and no hint that any such a mechanism should be at work in the composite
Higgs framework. The cutoff scale $\Lambda$ is always given by the strong sector scale
$m_*$, irregardless of the presence of accidentally light partners with mass
$m_{\textrm{light}}\ll m_*$. \footnote{In the extra-dimensional models $m_*$ is represented
by the compactification length, in the deconstructed ones it is provided by the fermonic
masses at the internal sites. We have checked explicitly that $\Lambda\simeq m_*$
in the deconstructed models presented in the following section.}
What lowers $m_H$ when the partners are light is not a lower cutoff but, more simply, a
smaller elementary/composite coupling. Indeed, by inverting eq.~(\ref{yyt}),
$y^2=y_tm_{\textrm{light}}/f$. By setting $\Lambda=m_*=g_*f_\pi$ one obtains again
eq.~(\ref{eq:mhgen}).
In conclusion, while the final formula is the same of ref.~\cite{Contino:2006qr}, the
derivation of the present section shows that it has a rather different physical origin.

\subsubsection*{Top mass in explicit models}

Before concluding this section we notice that the Lagrangian (\ref{lagt}) is significantly
more general than the one
we will actually encounter in the specific models. First of all, the concrete models
are more restrictive because they enjoy one more symmetry which has not
yet been taken into account in the discussion. This is ordinary \emph{parity}
invariance of the strong sector, which we always assume for simplicity
in our explicit constructions. Parity acts as
${\mathcal{O}}_L(\vec{x})\leftrightarrow{\mathcal{O}}^{(P)}_R(-\vec{x})$
on the operators in eq.~(\ref{mix}), and obviously it is broken by the interaction with the SM
particles. \footnote{The superscript ``$(P)$'' denotes the ordinary action
of parity on the Dirac spinors, for instance in the Weyl basis $\psi^{(P)}=\gamma^0\psi$}
However it can be formally restored by the method of spurions, we have to
assign transformations $q_L^{\mathbf5}(\vec{x})\leftrightarrow
{t_R^{\mathbf5}}^{(P)}(-\vec{x})$ to the embeddings, plus of course
$y_L\leftrightarrow y_R$. One implication of the parity symmetry is that the two
coefficients of the quadratic potential (\ref{pot2}) have to be equal,
$c_L=c_R$, and thus the relation among the $y_L$ and $y_R$ mixings (\ref{rel})
becomes simply $y_L\simeq\sqrt{2}y_R$. For what concerns instead the
partners Lagrangian (\ref{lagt}) parity implies $a_L=a_R$ and $b_L=b_R$.

Moreover, the additional symmetry structures which underly the formulation
of our models require the relations $a_L=a_R$ and $b_L=b_R$. The reason will become
more clear in the following section, the basic point is that in our
construction the fourplet and singlet form a fiveplet under an additional ${\textrm{SO}}(5)$
group which  is respected by the mixings.

To make contact with our models, let us then choose $a_L=a_R=b_L=b_R=1$, the top mass becomes
\begin{equation}
m_t \simeq \frac{| m^*_T -  m^*_{\widetilde{T}}|}{2 \sqrt{2}}\sin \varphi_L \sin \varphi_R
\sin\left(2 v/{f_\pi}\right)\,,\;\;\;\;\;{\textrm{with}}\;\;\;\;\;\;\;\;\left\{
\begin{array}{l}\sin\varphi_{L}=\frac{y_L f_\pi}{m_T}\\
\sin\varphi_{R}=\frac{y_R f_\pi}{m_{\widetilde{T}}}
\end{array}\right.\,,
\label{mt}
\end{equation}
and it is proportional to the mass-difference $m^*_T -  m^*_{\widetilde{T}}$. Indeed for  $a_L=a_R=b_L=b_R$
 the mixings are proportional to the five-plet $\Psi$ defined as
\begin{equation}
\Psi_I\,=\,U_{Ii}Q^i\,+\,U_{I5}{\widetilde{T}}\,=\,U_{I\bar{I}}\left(\begin{array}{c}Q\\{\widetilde{T}}\end{array}\right)_{\bar{I}}\,,
\end{equation}
which is related to the original fields by the orthogonal matrix $U$.
It becomes therefore convenient to perform a field redefinition and to re-express the Lagrangian in
terms of $\Psi$, in this way the mixings become trivial and independent of the Higgs field
and the only operators which contain the Higgs boson and no derivatives originate from the rotation of the
mass terms. Therefore these operators are proportional to the mass difference $m^*_{T}-m^*_{{\widetilde{T}}}$ because
for $m^*_{T}=m^*_{{\widetilde{T}}}$ also the mass Lagrangian becomes  ${\textrm{SO}}(5)$ invariant and
the dependence on the Higgs drops. Explicitly, we have
\begin{equation}
-{\overline{\Psi}} \,U \left(\begin{array}{cc}m_T^*&0\\0&m_{\widetilde{T}}^*\end{array}\right)U^t\Psi\,=\,
-m_T^*\overline{T}T-m_{\widetilde T}^*\overline{\widetilde{T}}\widetilde{T}-\,\frac{m_T^*-m_{\widetilde{T}}^*}{2\sqrt{2}}\sin(2h/f_\pi)\overline{T}\widetilde{T}\,+\,\ldots
\end{equation}
from which eq.~(\ref{mt}) is immediately rederived.

\section{Light partners in the DCHM$_\mathbf{3}$}

The first explicit model we will consider for our analysis is the $3$-site Discrete
Composite Higgs Model (DCHM$_3$)~\cite{Panico:2011pw}. This model provides
a simple but complete four-dimensional realization of the composite Higgs paradigm.
As we already mentioned in the Introduction, an important, distinctive property
of the DCHM$_3$ model is the finiteness and calculability of the Higgs potential.
This feature, together with the \emph{simplicity} of the DCHM approach,  will enable us
to derive explicit formulas displaying the relation between
the Higgs mass and the spectrum of the top partners.

Another important aspect is the fact that
the parametrization which we naturally get in the Discrete Composite Higgs
framework can be directly mapped onto the general structure of
partial compositeness. As we already showed in the previous section
partial compositeness plays a crucial role in understanding the relation
between the properties of the Higgs boson and the spectum of the fermionic resonances.
We will confirm this in the explicit analysis we will present
in this section.

\subsection{Structure of the model}

\begin{figure}[t]
\centering
\includegraphics[width=.65\textwidth]{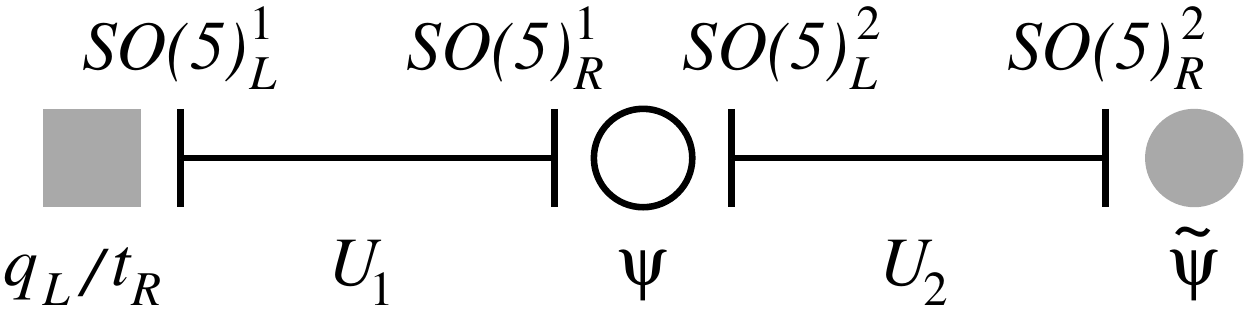}
\caption{Schematic structure of the three-site DCHM.}\label{fig:3-site}
\end{figure}

The basic structure of the DCHM$_3$ model consists of two replicas of the
non-linear $\sigma$-model \mbox{${\textrm {SO}}(5)_L \times  {\textrm {SO}}(5)_R/{\textrm {SO}}(5)_V$}.
The symmetry structure can be directly connected to a three-site pattern, as schematically shown in
figure~\ref{fig:3-site}, where each $\sigma$-model, whose degrees of freedom
are denoted by $U_{1,2}$, is represented by a link connecting two sites.
In this way we can relate each site to a corresponding subgroup of the
global invariance of the model.
In order to accommodate the hypercharges of the fermionic sector, an extra
${\textrm {U}}(1)_X$ global factor must be introduced. For simplicity, we do not associate this
Abelian factor to any of the sites and let it act on all the fermions of the
model. \footnote{For a more detailed discussion of this point see footnote~9
of Ref.~\cite{Panico:2011pw}.}

The elementary gauge fields, as well as the vector resonances coming from the composite
sector, are introduced by gauging suitable subgroups of the total global invariance
$({\textrm {SO}}(5))^4\times {\textrm {U}}(1)_X$. The elementary fields correspond to the gauging of
an ${\textrm {SU(2)}}_L \times {\textrm {U}}(1)_Y$ subgroup of ${\textrm {SO}}(5)_L^1 \times {\textrm {U}}(1)_X$,
with the identification of the hypercharge $Y = T^3_R +X$. Two levels of composite resonances
are introduced by gauging at the middle and last site. At the middle site we
gauge the diagonal subgroup ${\textrm {SO}}(5)_D$ of the global invariance ${\textrm {SO}}(5)^1_R \times {\textrm {SO}}(5)^2_L$.
At the last site we gauge only an ${\textrm {SO}}(4)$ subgroup of the ${\textrm {SO}}(5)^2_R$ global invariance.
The model encodes, through the explicit breaking induced by the gauging at the last site, the spontaneous
${\textrm {SO}}(5) \rightarrow {\textrm {SO}}(4)$ global symmetry breaking pattern of the strong sector.

A useful form of the Lagrangian, which is suitable for the computation of the spectrum
and of the Higgs potential, is obtained by adopting the ``holographic'' gauge. \footnote{This
terminology is inspired from the extra-dimensional holographic technique~\cite{Panico:2007qd}.}
In this gauge, the only dependence on the Goldstone degrees of freedom
appears at the first site, where the elementary fields live, while the Lagrangian of the composite
sector assumes a particularly simple form. As explained in~\cite{Panico:2011pw}, the holographic gauge
can be reached by gauge transformations at the middle and last site which set $U_2$ equal to the identity and
remove the unphysical degrees of freedom from $U_1$. After the transformation $U_1$ becomes the Goldstone
matrix $U$ in eq.~(\ref{go}).

The fermionic sector of the model contains the elementary fields corresponding to the SM
chiral fermions and two sets of composite resonances. For simplicity we will only
focus on the third quark generation and in particular on the fields related to the top
quark, and we will neglect the light generations, the right-handed bottom field
$b_R$ and the corresponding composite partners. This approximation is justified by the fact
that the contribution of the latter to the Higgs effective potential is parametrically
suppressed by the small bottom mass with respect to the one coming from the top tower.

In analogy with the gauge sector, the composite fermionic resonances are introduced
at the middle and the last site. At the middle site we add one multiplet $\psi$,
which transforms in the fundamental representation of the vector group $\textrm{SO}(5)_D$
and has $\textrm{U}(1)_X$ charge $2/3$.
Another multiplet in the fundamental representation of $\textrm{SO}(5)_R^2$ is introduced
at the last site $\widetilde \psi \in {\bf 5}_{2/3}$.
Given that, at the last site, the $\textrm{SO}(5)_R^2$ invariance is broken,
it is useful to introduce also a notation for the decomposition of the $\widetilde \psi$
multiplet in representations of the unbroken
$\textrm{SO}(4) \simeq \textrm{SU}(2)_L \times \textrm{SU}(2)_R$ subgroup.
The fundamental representation of $\textrm{SO}(5)$ decomposes as
${\bf 5} = ({\bf 2}, {\bf 2}) \oplus ({\bf 1}, {\bf 1})$, thus
\begin{equation}
\widetilde \psi =
\left(
\begin{array}{c}
\widetilde Q\\
\widetilde T
\end{array}
\right)\,,
\end{equation}
where $\widetilde Q \in ({\bf 2}, {\bf 2})$, while $\widetilde T$ is a singlet.
The Lagrangian for the composite states $\psi$ and $\widetilde \psi$
in the holographic gauge is given by
\begin{eqnarray}
{\cal L}_{\textrm{comp}}^{\textrm{f}} &=& i\, \overline \psi \Dslash \psi - m \overline \psi \psi\nonumber\\
&& +\, i\, \overline{\widetilde\psi} \Dslash \widetilde\psi
- \widetilde m_{\textrm{Q}} \overline{\widetilde Q} {\widetilde Q}
- \widetilde m_{\textrm{T}} \overline{\widetilde T} \widetilde T\nonumber\\
&& -\, \Delta \overline \psi \widetilde \psi + {\rm h.c.}\,.
\label{eq:lag_ferm_holo}
\end{eqnarray}
In the above expression we included a breaking of the $\textrm{SO}(5)_R^2$ group
through the mass terms for $\widetilde Q$ and $\widetilde T$, which preserve only
the $\textrm{SO}(4)$ subgroup. Notice that the mixing on the last line of
eq.~(\ref{eq:lag_ferm_holo}) comes from a term of the form
$\Delta\, \overline \psi\, U_2\, \widetilde \psi + {\rm h.c.}$,
which appears in the original non-gauge-fixed Lagrangian.

The elementary fermions are introduced at the first site. They are given by the SM
chiral states $q_L$ and $t_R$. The terms in the Lagrangian which involve the elementary
fermions are given by \footnote{In the Lagrangian ${\cal L}^{\textrm{f}}_{\textrm{elem}}$
we use a different normalization of the left mixing $y_L$ with respect
to the choice in the corresponding eq.~(54) of Ref.~\cite{Panico:2011pw}.\label{foot:yL}}
\begin{equation}
{\cal L}^{\textrm{f}}_{\textrm{elem}} = i\, \overline q_L \Dslash q_L
+ i\, \overline t_R \Dslash t_R
-\, y_L f \overline q_L^{\bf 5} U \psi_{R}
- y_R\, f \overline t_R^{\bf 5} U \psi_{L} + {\rm h.c.}\,,
\label{eq:elem_lag_holo}
\end{equation}
where we used the embeddings of the elementary states in the fundamental representation
of ${\textrm {SO}}(5)$ given in eq.~(\ref{emb}). Following the notation of~\cite{Panico:2011pw},
we write the elementary--composite mixings in terms of the Goldstone decay
constant $f$ of the two fundamental ${\textrm {SO}}(5)_L \times {\textrm {SO}}(5)_R$ non-linear $\sigma$-models.
This quantity is related to the Higgs decay constant by $f_\pi = f/\sqrt{2}$.

\subsection{The Higgs potential}

In this section we will analyze the structure of the Higgs potential
deriving an approximate expression for the Higgs mass in terms of the masses
of the fermionic resonances.

The most relevant contribution to the Higgs potential comes from the
fermionic states. The corrections due to the gauge fields are typically small
and we will neglect them altogether in our analysis. The only states
which are coupled to the Higgs in our set-up are the top and the resonances of charge $2/3$.
The contribution of these states to the potential has the form \footnote{The computation
of the Higgs potential can be performed by using the standard textbook formulae for the
Coleman--Weinberg potential. Equivalently one can apply the holographic technique
as explained in Ref.~\cite{Panico:2007qd}.}
\begin{equation}\label{eq:potential}
V(h) = -\frac{2 N_c}{8 \pi^2} \int d p\, p^3
\log\left(1 - \frac{C_1(p^2) \sin^2 (h/f_\pi) + C_2(p^2) \sin^2 (h/f_\pi) \cos^2 (h/f_\pi)}{D(p^2)}\right)\,.
\end{equation}
The denominator of the expression in the logarithm is given by
\begin{equation}\label{eq:denom}
D(p^2) = 2 p^2 \prod_{I=T,{\widetilde T},T_{2/3}}\big(p^2 + m_{I_-}^2\big)\big(p^2 + m_{I_+}^2\big)\,,
\end{equation}
where $m_{I_\pm}$ denote the masses of the charge $2/3$ resonances before EWSB.
In particular $T$ and $T_{2/3}$ denote the two states in the fourplet,
namely $T$ is the state which forms an $SU(2)_L$ doublet with the charge $-1/3$
field ($B$) and $T_{2/3}$ is the state which appear in the doublet with the exotic
state of charge $5/3$ ($X_{5/3}$). The $\widetilde T$ state denotes instead the singlet.
The $\pm$ sign refers to the two levels of composite resonances which are present in the model.
Notice that all these masses include the shift due to the mixings with the elementary
states. The initial factor $p^2$ which appears in eq.~(\ref{eq:denom})
is due to the presence of the top which is massless before EWSB.
The coefficients appearing in eq.~(\ref{eq:denom}) in the numerator of the expression inside the
logarithm are given by
\begin{equation}
\left\{
\begin{array}{l}
C_1(p^2) = \left(y_L^2 - 2 y_R^2\right) f^2 F_1(p^2) F_2(p^2) - (\widetilde m_Q - \widetilde m_T) \Delta^2
y_L^2 y_R^2 f^4\left(p^2 + \Delta^2 + \widetilde m_Q^2\right) F_1(p^2)\\
\rule{0pt}{1.5em}C_2(p^2) = -(\widetilde m_Q - \widetilde m_T) \Delta^2 y_L^2 y_R^2 f^4 F_2(p^2)
\end{array}
\right.\,,
\end{equation}
where the functions $F_{1,2}$ are defined as
\begin{equation}
\left\{
\begin{array}{l}
F_1(p^2) = p^2 \Big((m + \widetilde m_T)(p^2 + \Delta^2 - m\, \widetilde m_Q)
+ (m + \widetilde m_Q)(p^2 + \Delta^2 - m\, \widetilde m_T) \Big)\\
\rule{0pt}{1.5em}F_2(p^2) = (\widetilde m_Q - \widetilde m_T) \Delta^2
\left(p^2 + m_{{T_{2/3-}}}^2\right) \left(p^2 + m_{{T_{2/3+}}}^2\right)
\end{array}
\right.\,.
\end{equation}

The potential can be approximated by expanding at leading order the logarithm in
eq.~(\ref{eq:potential}). Although this approximation is formally valid only for small
values of $h/f_\pi$, \footnote{This is of course
not true in the limit $p^2 \rightarrow 0$, in which the argument of the logarithm
diverges. However in this case the factor $p^3$ in front of the
logarithm compensate for the divergence and the approximate integrand vanishes for
$p \rightarrow 0$. The error introduced by this approximation is thus small.}
it turns out that it is numerically very accurate in a wide range of the parameter
space and, in particular, it is valid for all the points we will consider in our
numerical analysis.

After the expansion and the integration, the potential takes the general form
already considered in eq.~(\ref{tun0})
\begin{equation}\label{eq:pot_expansion}
V(h) \simeq \alpha \sin^2 (h/f_\pi) - \beta \sin^2 (h/f_\pi) \cos^2 (h/f_\pi)\,.
\end{equation}
Using an expansion in the elementary mixings,
the $\alpha$ term is dominated by the leading ${\cal O}(y^2)$ contributions,
proportional to $y_L^2 - 2 y_R^2$. As discussed in section~\ref{sec:genan}, in order
to obtain a realistic value for $v/f_\pi$ the leading order contributions
must be cancelled, such that they can be tuned against
the subleading terms. This leads to the condition in eq.~(\ref{rel})
with $c_L = c_R = 1$, namely
\begin{equation}\label{eq:initun}
y_L \simeq \sqrt{2} y_R\,.
\end{equation}
This relation is very well verified numerically for realistic points in the
parameter space, as shown in~\cite{Panico:2011pw}. \footnote{The condition
in eq.~(\ref{eq:initun}) differs from the one reported in eq.~(57)
of~\cite{Panico:2011pw} by a factor $\sqrt{2}$. This is due to a different choice of the
normalization of the $y_L$ mixing (see eq.~(\ref{eq:elem_lag_holo}) and footnote~\ref{foot:yL}).}

For realistic configurations, due to the cancellation, the leading term of order $y_{L,R}^2$
becomes of ${\cal O}(y_{L,R}^4)$. This means that, if we are interested in an
expansion of the potential at quartic order in the elementary--composite
mixings, we only need to take the linear term in the expansion of the
logarithm in eq.~(\ref{eq:potential}).
The value of the coefficient $\beta$ can be easily found analytically
\begin{equation}\label{eq:exact_B}
\beta = \frac{N_c}{8 \pi^2}(\widetilde m_Q - \widetilde m_T)^2\, \Delta^4\, y_L^2\, y_R^2\, f^4
\!\!\!\!\sum_{\begin{array}{l@{}c@{}l}\vspace{-.25em}\scriptstyle I &\scriptstyle =& \scriptstyle T_-\!, T_+\!,\\
&&\scriptstyle {\widetilde T}_-\!, {\widetilde T}_+\end{array}}
\frac{\log(m_I/f)}{\prod_{J \neq I} (m_I^2 - m_J^2)}\,.
\end{equation}
In the limit in which the second level of resonances is much heavier than the first one,
we can use an expansion in the ratio of the heavy and light states masses and get a simple
approximate formula for $\beta$:
\begin{equation}\label{eq:approx_B}
\beta \simeq \frac{N_c}{8 \pi^2}(\widetilde m_Q - \widetilde m_T)^2\, \Delta^4\, y_L^2\, y_R^2\, f^4
\frac{\log\big(m_{T_-}/m_{{\widetilde T}_-}\big)}{\big(m_{T_-}^2 - m_{{\widetilde T}_-}^2\big)
m_{T_+}^2 m_{{\widetilde T}_+}^2}\,.
\end{equation}
As can be seen form this formula, when one of the states $T_-$ or $\widetilde T_-$
is much lighter than the other, the contribution to $\beta$ from the first level of resonances
is enhanced by the logarithmic factor $\log(m_{T_-}/m_{{\widetilde T}_-})$.
In this case the light states contribution completely dominates and the
corrections due to the second layer of resonances become negligible.
On the other hand, if the two light states have comparable masses, the second level of
resonances, in certain regions of the parameter space, can be relatively close in mass to the first one,
thus giving sizable corrections to the Higgs mass. The sign of these corrections is fixed,
and they always imply a decrease of the Higgs mass. The size of the corrections in the relevant
regions of the parameter space is typically below $50\%$.

The expression of the Higgs mass in terms of the $\beta$ coefficient has already been
given in eq.~(\ref{mh}) and reads
\begin{equation}\label{eq:mh_appB}
m_H^2 = \frac{2 \beta}{f_\pi^2} \sin^2(2v/f_\pi)
\simeq \frac{N_c}{\pi^2}(\widetilde m_Q - \widetilde m_T)^2\, \Delta^4\, y_L^2\, y_R^2\, f_\pi^3
\frac{\log\big(m_{T_-}/m_{{\widetilde T}_-}\big)}{\big(m_{T_-}^2 - m_{{\widetilde T}_-}^2\big)
m_{T_+}^2 m_{{\widetilde T}_+}^2} \sin^2(2v/f_\pi)\,.
\end{equation}

\subsection{The Higgs mass and the top partners}

As shown in the general analysis of section~\ref{sec:higgspartners}, it
is useful to compare the Higgs mass with the top mass, with the aim of obtaining
a relation between $m_h$ and the masses of the top partners.

By performing an expansion in $\sin^2(v/f_\pi)$, we can obtain an approximate
expression for the top mass. The result can be recast in the general form
of eq.~(\ref{mt}),
\begin{equation}\label{eq:app_top_mass_1}
m_t \simeq \frac{|{\widetilde m}_Q - {\widetilde m}_T|}{2 \sqrt{2}}
\sin \varphi_L \sin \varphi_R
\sin\left(\frac{2 v}{f_\pi}\right)\,.
\end{equation}
where the mixing angles $\varphi_{L,R}$ are now replaced by some ``effective'' compositeness angles
\begin{equation}\label{eq:effcomp}
\begin{array}{l}
\sin \varphi_L \equiv \displaystyle
\frac{\Delta}{\sqrt{\Delta^2 + {\widetilde m}_Q^2}}
\frac{y_L f}{\sqrt{\frac{(\Delta^2 - m\, \widetilde m_Q)^2}{\Delta^2 + {\widetilde m}_Q^2} + (y_L f)^2}}\,,\\
\sin \varphi_R \equiv \displaystyle
\frac{\Delta}{\sqrt{\Delta^2 + {\widetilde m}_T^2}}
\frac{y_R f}{\sqrt{\frac{(\Delta^2 - m\, \widetilde m_T)^2}
{\Delta^2 + {\widetilde m}_T^2} + (y_R f)^2}}\,.
\end{array}
\end{equation}

There is an equivalent way to rewrite the approximate expression for the top
mass in eq.~(\ref{eq:app_top_mass_1}) in terms of the masses of the
$T$ and $\widetilde T$ resonances:
\begin{equation}\label{app_t_mass_2}
m_t \simeq \frac{|{\widetilde m}_Q - {\widetilde m}_T|}{2 \sqrt{2}}
\frac{y_L y_R f^2 \Delta^2}{m_{T_+} m_{T_-} m_{{\widetilde T}_+} m_{{\widetilde T}_-}}
\sin\left(\frac{2 v}{f_\pi}\right)\,.
\end{equation}
By comparing this expression with the approximate formula for the Higgs mass
in eq.~(\ref{eq:mh_appB}) we find a remarkable relation
between $m_h$ and the masses of the lightest $T$ and $\widetilde T$
resonances:
\begin{equation}\label{eq:mHmt}
\frac{m_H}{m_t} \simeq \frac{\sqrt{2 N_c}}{\pi} \frac{m_{T_-} m_{{\widetilde T}_-}}{f_\pi}
\sqrt{\frac{\log\big(m_{T_-}/m_{{\widetilde T}_-}\big)}{m_{T_-}^2 - m_{{\widetilde T}_-}^2}}\,.
\end{equation}
As discussed in the previous section, the above expression receives
the corrections due to the presence of the second layer of resonances.
These corrections are sizable only when the second level of resonances
is relatively light. In this case corrections of the order $50\%$ to
eq.~(\ref{eq:mHmt}) can arise.

Let us now compare the expression in eq.~(\ref{eq:mHmt}) with the general
result obtained in section~\ref{sec:genan} (eq.~(\ref{eq:mhgen})).
The two equations show the same qualitative relation between the Higgs
mass and the masses of the lightest resonances $T$ and $\widetilde T$.
In the case $m_T = m_{\widetilde T}$ the two expressions exactly coincide, while,
when a large hierarchy between the two light states $T$ and $\widetilde T$
is present, they differ by a coefficient of ${\cal O}(1)$.
This shows that the general analysis of section~\ref{sec:genan} correctly
capture the main connection between the Higgs and the top partners masses,
both at a qualitative and a quantitative level. Notice that also the logarithmic term, which
originates from the one in the Higgs mass (\ref{eq:mh_appB}), could have been computed
within the general approach of section~\ref{sec:genan}. It is indeed an IR loop effect associated to
the light top partners.

\begin{figure}[t]
\centering
\includegraphics[width=.45\textwidth]{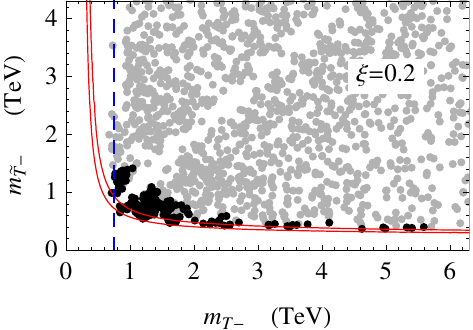}
\hspace{1.5em}
\includegraphics[width=.45\textwidth]{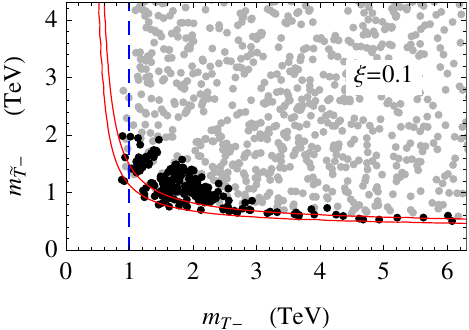}
\caption{Scatter plots of the masses of the lightest $T$ and $\widetilde T$ resonances
for $\xi = 0.2$ (left panel) and $\xi = 0.1$ (right panel) in the three-site DCHM model. The black dots
denote the points for which $115\ {\rm GeV} \leq m_H \leq 130\ {\rm GeV}$, while
the gray dots have $m_H > 130\ {\rm GeV}$. The scans have been obtained by
varying all the composite sector masses in the range $[-8 f, 8f]$ and keeping
the top mass fixed at the value $m_t = 150\ {\rm GeV}$.
The area between the solid red lines represents the range obtained by
applying the result in eq.~(\ref{eq:mHmt}) for $115\ {\rm GeV} \leq m_H \leq 130\ {\rm GeV}$.
The dashed blue line corresponds to the estimate of the lower bound on
$m_{T-}$ given in eq.~(\ref{eq:mTlow}).}\label{fig:mBid_mSing_DCHM3}
\end{figure}

\begin{figure}[t]
\centering
\includegraphics[width=.45\textwidth]{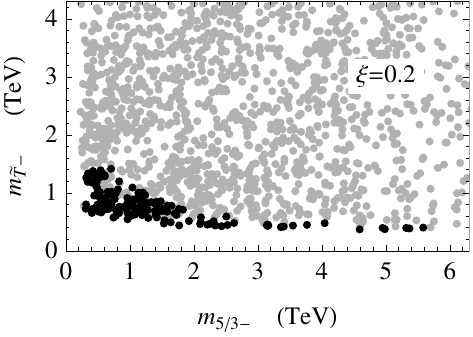}
\hspace{1.5em}
\includegraphics[width=.45\textwidth]{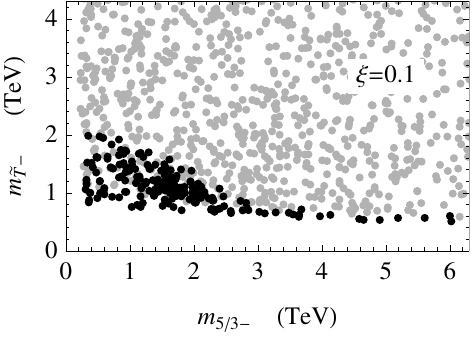}
\caption{Scatter plots of the masses of the lightest exotic state of charge $5/3$
and of the lightest $\widetilde T$ resonance for $\xi = 0.2$ (left panel) and $\xi = 0.1$ (right panel)
in the three-site DCHM model.
The black dots denote the points for which $115\ {\rm GeV} \leq m_H \leq 130\ {\rm GeV}$, while
the gray dots have $m_H > 130\ {\rm GeV}$. The scans have been obtained by
varying all the composite sector masses in the range $[-8 f, 8f]$ and keeping
the top mass fixed at the value $m_t = 150\ {\rm GeV}$.}\label{fig:mEx_mSing_DCHM3}
\end{figure}

We checked numerically the validity of our results by a scan on the parameter space
of the model.
In our numerical analysis we take the interval $115\ {\rm GeV} \leq m_H \leq 130\ {\rm GeV}$
as the range of Higgs masses compatible with the current LHC exclusion bounds.
This range has been chosen slightly larger than the current exclusion for a SM-like Higgs
to take into account the corrections due to the composite nature of the Higgs
\cite{Espinosa:2010vn}. In our analysis we also fix the top mass
to the value $m_t = m_t^{\overline{MS}}(2\ {\rm TeV}) = 150\ {\rm GeV}$,
which corresponds to $m_t^{pole} = 173\ {\rm GeV}$.

The scatter plots of the masses of the $T$ and $\widetilde T$
light resonances are shown in fig.~\ref{fig:mBid_mSing_DCHM3}.
One can see that eq.~(\ref{eq:mHmt}) describes accurately the relation between the
Higgs and the resonance masses in the regions in which one state is significantly
lighter than the others. For a realistic Higgs mass
this happens only when the $\widetilde T_-$ is much lighter than the other states.
Instead, the situation of a $T$ much lighter than the $\widetilde T$ can not happen
for a light Higgs due to the presence of a lower bound on the $m_{T_-}$, which will be
discussed in details in the next section.
In the region of comparable $T_-$ and $\widetilde T_-$ masses sizable deviations from
eq.~(\ref{eq:mHmt}) can occur. These are due to the possible presence of a
relatively light second level of resonances, as already discussed.

The numerical results clearly show that resonances with a mass of the order or below
$1.5\ {\rm TeV}$ are needed in order to get a realistic Higgs mass both in the case
$\xi = 0.2$ and $\xi = 0.1$. The prediction is even sharper for the cases in which only one
state, namely the $\widetilde T_-$, is light. In these regions of the parameter space
a light Higgs requires states with masses around $400\ {\rm GeV}$ for the $\xi = 0.2$ case
and around $600\ {\rm GeV}$ for $\xi = 0.1$.

The situation becomes even more interesting if we also consider the masses of the other composite
resonances. As we already discussed, the first level of resonances contains, in addition to the
$T_-$ and $\widetilde T_-$, three other states: a top-like state, the $T_{2/3-}$, a bottom-like
state, the $B_-$, and an exotic state with charge $5/3$, the $X_{5/3-}$. These three states together
with the $T_-$ form a fourplet of ${\textrm{SO(4)}}$. Obviously the $X_{5/3-}$ cannot mix with any other
state even after EWSB, and therefore it remains always lighter than the other particles in the fourplet.
In particular (see fig.~\ref{fig:Mass_spectrum} for a schematic
picture of the spectrum), it is significantly lighter than the $T_-$ .
In fig.~\ref{fig:mEx_mSing_DCHM3} we show the scatter plots of the masses of the lightest exotic
charge $5/3$ state and of the $\widetilde T$.
In the parameter space region in which the Higgs is light the $X_{5/3-}$ resonance can be much
lighter than the other resonances, especially in the configurations in which the $T_-$
and $\widetilde T_-$ have comparable masses.
In these points the mass of the exotic state can be as low as $300\ {\rm GeV}$.

Notice that in the plots in fig.~\ref{fig:mBid_mSing_DCHM3} there are no points in which the
masses of the $T_-$ and of the $\widetilde T_-$ coincide.
This is due to a repulsion of the mass levels induced by the mixings due to EWSB.
As expected, this effect is more pronounced for larger values of $\xi$.

\subsection{The top mass and a lower bound on the Higgs mass}

As  noticed above, the asymptotic
region $m_{T-} \ll m_{\widetilde T-}$, which could in principle give rise to configurations
with realistic Higgs masses, is not accessible in our model. Indeed in the scatter plots of
fig.~\ref{fig:mBid_mSing_DCHM3} we find a lower bound on $m_{T-}$. We will show below that
this bound comes from the requirement of obtaining a realistic top mass and that an analogous bound,
which however is not visible in fig.~\ref{fig:mBid_mSing_DCHM3}, exists for the $\widetilde T_-$ mass.
From these results we will also derive an absolute lower bound on the Higgs mass.

The starting point of our analysis is the approximate expression
for the top mass in eq.~(\ref{eq:app_top_mass_1}). Our aim is to abtain a lower bound
on the resonance masses, so we will focus on the configurations in which one of
the top partners is much lighter than the others. For definiteness we will
consider the case in which the lightest state is the $T_-$ resonance.
In a generic situation, all the parameters of the composite sector are of the same order
$\Delta \sim m \sim \widetilde m_Q \sim \widetilde m_T$. The only mass which gets
cancelled is $m_{T_-}$, so we can also assume that $m_{T_+} \sim m_{\widetilde T_+}$
and that they are of the same order of the composite sector masses.
In this regime the effective compositeness angles in eq.~(\ref{eq:effcomp})
can be approximated as
\begin{equation}
\sin \varphi_L \sim 1\,,
\qquad
\sin \varphi_R \simeq \frac{y_R f}{m_{\widetilde T_-}}\,.
\end{equation}
The first equation comes from the fact that we assumed the $T_-$ state to be nearly
massless before the mixing with the elementary sector. This condition is equivalent
to the relation $\Delta^2 - m\, \widetilde m_Q = 0$ (see eq.~(80) of \cite{Panico:2011pw}).

The expression for the top mass in eq.~(\ref{eq:app_top_mass_1}) now becomes
\begin{equation}
m_t \simeq \frac{y_R f}{2 \sqrt{2}} \sin\left(\frac{2 v}{f_\pi}\right) \simeq y_R v\,,
\end{equation}
and, by using the relation between $y_L$ and $y_R$ in eq~(\ref{eq:initun}), we get
\begin{equation}
y_L \simeq \sqrt{2}\, y_R \simeq \frac{\sqrt{2} m_t}{v}\,.
\end{equation}
Given that the mass of the light state predominantly comes from the mixing with the
elementary fermions we can use the estimate
\begin{equation}\label{eq:mTlow}
m_{T_-} \gtrsim y_L f \simeq \frac{2 m_t}{v} f_\pi\,.
\end{equation}
This inequality implies the lower bounds
\begin{equation}
m_{T_-} \gtrsim 5 m_t \simeq 750\ {\rm GeV}\,, \qquad {\rm for}\ \ \xi = 0.2\,,
\end{equation}
and
\begin{equation}
m_{T_-} \gtrsim 6.7 m_t \simeq 1000\ {\rm GeV}\,, \qquad {\rm for}\ \ \xi = 0.1\,,
\end{equation}
obtained for $m_t = 150\ {\rm GeV}$.
In a similar way a lower bound on the mass of the lightest $\widetilde T$ state
can be found. This bound is a factor $2$ weaker than the one on $m_{T-}$:
\begin{equation}
m_{\widetilde T_-} \gtrsim y_R f \simeq \frac{m_t}{v} f_\pi\,.
\end{equation}
The lower bounds on the lightest top partners masses agree with the results
of the numerical scans in fig.~\ref{fig:mBid_mSing_DCHM3}. The lower bound on
$m_{\widetilde T_-}$ is instead below the range of values needed to get a realistic Higgs mass,
so it is not visible in the the plot.

The lower bound on the resonance masses can be translated, through
eq.~(\ref{eq:mHmt}) into a lower bound on the Higgs mass. The most favourable
configuration is the one in which the lightest mass is $m_{\widetilde T_-}$.
This leads to the bound
\begin{equation}
m_H \gtrsim \frac{\sqrt{2 N_c}}{\pi}\frac{m_t^2}{v}\sqrt{\log\left(\frac{v}{m_t}\frac{m_{T_-}}{f_\pi}\right)}\,.
\end{equation}
For $m_{T_-}/f_\pi \sim 4$, which represent a typical point in our parameter space, we get
\begin{equation}
m_H \gtrsim 100\ {\rm GeV}\,.
\end{equation}
This result is in good agreement with the bound obtained in the scans.

\section{The simplest composite Higgs model}

As shown in Ref.~\cite{Panico:2011pw}, the three-site DCHM we considered in the previous section
is the minimal realization of an effective description of a composite Higgs in which
all the key observables, and in particular the Higgs potential, are computable
at the leading order. This property allowed us to decouple the UV physics and
fully characterize the model in terms of the parameters describing the elementary
states and two levels of composite resonances.

\begin{figure}[t]
\centering
\includegraphics[width=.35\textwidth]{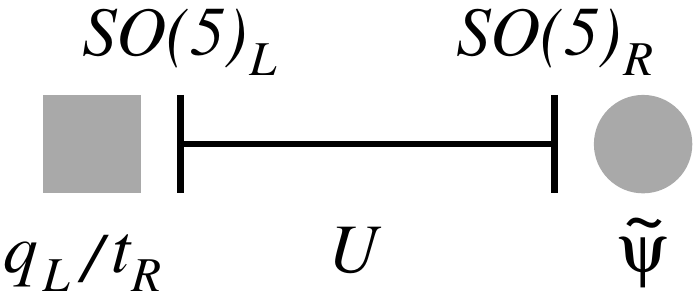}
\caption{Schematic structure of the two-site DCHM.}\label{fig:2-site}
\end{figure}
If we accept to give up a complete predictivity, a much simpler effective model can be
employed to describe the low-energy dynamics of a composite Higgs boson and of the top partners.
In this model only
one layer of composite resonances is introduced, leading to a structure representable
with a two-site model (see fig.~\ref{fig:2-site}).
The pattern of divergences in the two-site DCHM has been fully analyzed in Ref.~\cite{Panico:2011pw}:
the electroweak precision parameters remain calculable at leading order, while the
Higgs potential becomes logarithmically divergent at one loop.

There is however an interesting property which partially preserves predictivity
also for the potential. In the expansion in powers of the elementary--composite
mixings, only the leading terms can develop a logarithmic divergence, while the higher order ones
are finite at one loop.  We have shown in Section~2.1 (see eq.~(\ref{eq:oper})) that at the leading order only
two operators exist and that they both give the same contribution, proportional to $\sin^2{h/f_\pi}$, to the
potential. A single counterterm is therefore enough to regulate the divergence, which corresponds to the
renormalization of a single parameter.
An interesting possibility is to fix the value of the Higgs VEV, or more precisely of the ratio $v/f_\pi$,
as renormalization condition obtaining the Higgs mass as a prediction. In this sense, $m_H$ is predictable
also in the DCHM$_2$.

\subsection{Structure of the 2-site model}

Let us briefly summarize the structure of the DCHM$_2$.
The model is based on a non-linear $\sigma$-model ${\textrm {SO}}(5)_L \times {\textrm {SO}}(5)_R/{\textrm {SO}}(5)_V$ and
it is schematically representation in fig.~\ref{fig:2-site}.
As in the three-site DCHM, the first site is associated with the elementary states,
while the other is related to the composite resonances. Of course, in this case,
only one level of composite resonances is present. In order to accommodate
the hypercharge for the fermions an extra ${\textrm {U}}(1)_X$ symmetry
must be introduced, which acts on the fermion fields at both sites.

The elementary gauge bosons are added at the first site by gauging an $SU(2)_L \times {\textrm {U}}(1)_Y$
subgroup of ${\textrm {SO}}(5)_L \times {\textrm {U}}(1)_X$, with the choice of the hypercharge
as $Y = T_R^3 + X$. The composite gauge resonances are in the adjoint representation
of ${\textrm {SO}}(4)$ and gauge a corresponding subgroup of ${\textrm {SO}}(5)_R$.

One level of composite fermions $\widetilde \psi$ is introduced at the second site.
They transform in the fundamental representation of the ${\textrm {SO}}(5)_R$ global group
and have ${\textrm {U}}(1)_X$ charge $2/3$. Analogously to the three-site case,
the spontaneous breaking of ${\textrm {SO}}(5)$ in the composite
sector is parametrized by the explicit breaking of the additional ${\textrm {SO}}(5)_R$ global group.
In the fermionic sector this is achieved by a mass term which only respects the ${\textrm {SO}}(4)$
subgroup. The Lagrangian for the composite states $\widetilde \psi$, in the holographic gauge, is given by
\begin{equation}
{\cal L}_{\textrm{comp}}^{\textrm{f}}
= i\, \overline{\widetilde\psi} \Dslash \widetilde\psi
- \widetilde m_{\textrm{Q}} \overline{\widetilde Q} {\widetilde Q}
- \widetilde m_{\textrm{T}} \overline{\widetilde T} \widetilde T\,,
\label{eq:lag_ferm_holo_2-site}
\end{equation}
where we have split $\widetilde \psi$ in ${\textrm {SO}}(4)$ representation, ${\bf 5} = ({\bf 2}, {\bf 2}) \oplus ({\bf 1}, {\bf 1})$, as
\begin{equation}
\widetilde \psi =
\left(
\begin{array}{c}
\widetilde Q\\
\widetilde T
\end{array}
\right)\,,
\end{equation}
where $\widetilde Q \in ({\bf 2}, {\bf 2})$ and $\widetilde T$ is the singlet.

The elementary fermions, {\it{i.e.}} the SM chiral states $q_L$ and $t_R$, are introduced at the first site. Their
Lagrangian is
\begin{equation}
{\cal L}^{\textrm{f}}_{\textrm{elem}} = i\, \overline q_L \Dslash q_L
+ i\, \overline t_R \Dslash t_R
-\, y_L f_\pi \overline q_L^{\bf 5} U \widetilde \psi_{R}
- y_R\, f_\pi \overline t_R^{\bf 5} U \widetilde \psi_{L} + {\rm h.c.}\,,
\label{eq:elem_lag_holo_2-site}
\end{equation}
where we used the embeddings of the elementary states in the fundamental representation
of ${\textrm {SO}}(5)$ given in eq.~(\ref{emb}).

Notice that we have already encountered the fermion Lagrangian of the DCHM$_2$ in the general discussion
of section~2, and in particular at the end of section~\ref{sec:genan}. The DCHM$_2$ can indeed be obtained from
the general Lagrangian of eq.~(\ref{lagt}) by restricting $a_L=a_R=b_L=b_R$ in order to respect the ${\textrm {SO}}(5)$
symmetry.

\subsection{The Higgs potential}

Analogously to the DCHM$_3$ case, the fermionic contribution to the Higgs potential only
comes from the charge $2/3$ states. Its structure can be put in the same form as
eq.~(\ref{eq:potential})
\begin{equation}\label{eq:potential_2-site}
V(h) = -\frac{2 N_c}{8 \pi^2} \int d p\, p^3
\log\left(1 - \frac{C_1(p^2) \sin^2 (h/f_\pi) + C_2(p^2) \sin^2 (h/f_\pi) \cos^2 (h/f_\pi)}{D(p^2)}\right)\,.
\end{equation}
The denominator of the expression in the logarithm now contains only one level of resonances
and is given by
\begin{equation}
D(p^2) = 2 p^2 \prod_{I=T,{\widetilde T},T_{2/3}}\big(p^2 + m_{I}^2\big)\,,
\end{equation}
where we used a notation similar to the one adopted for the three-site model.
For the two-site model the expression for the masses of the top parteners before EWSB are
very simple and can be given in closed form
\begin{equation}
m_T^2 = \sqrt{\widetilde m_Q^2 + (y_L f_\pi)^2}\,, \qquad
m_{T_{2/3}}^2 = \widetilde m_Q^2\,, \qquad
m_{\widetilde T}^2 = \sqrt{\widetilde m_T^2 + (y_R f_\pi)^2}\,.
\end{equation}
The $C_{1,2}$ coefficients appearing in the expression of the Higgs potential are given by
\begin{equation}
\left\{
\begin{array}{l}
C_1(p^2) = -\left({\widetilde m}_Q^2 - {\widetilde m}_T^2\right)
p^2 \left(\left(p^2 + m_{T_{2/3}}^2\right) (y_L^2 - 2 y_R^2) f_\pi^2 - y_L^2 y_R^2 f_\pi^4\right)\\
\rule{0pt}{1.5em}C_2(p^2) = -({\widetilde m}_Q- {\widetilde m}_T)^2
\left(p^2 + m_{T_{2/3}}^2\right) y_L^2 y_R^2 f_\pi^4
\end{array}
\right.\,.
\end{equation}

Similarly to the three-site model, the second term appearing in the logarithm argument
in eq.~(\ref{eq:potential_2-site}) is typically much smaller than one, so that
we can use a series expansion. \footnote{For
more details see the discussion before eq.~(\ref{eq:pot_expansion}).}
The potential, taking into account terms up to the quartic order in the elementary--composite mixings,
has the usual form
\begin{equation}\label{eq:pot_expansion_2-site}
V(h) \simeq \alpha \sin^2 (h/f_\pi) - \beta \sin^2 (h/f_\pi) \cos^2 (h/f_\pi)\,.
\end{equation}
As we already mentioned, the ${\cal O}(y_{L,R}^2)$ terms in the potential are
logarithmically divergent, as can be easily checked using the explicit
results given above. This implies that the coefficient $\alpha$ in
eq.~(\ref{eq:pot_expansion_2-site}) must be regularized. For this purpose
we can add a counterterm of the form given in eq.~(\ref{eq:oper}) with
a suitable coefficient. This procedure is equivalent, from a practical point of view, to just
consider $\alpha$ as a free parameter. This coefficient can then be fixed by imposing
one renormalization condition, for instance by choosing the value of $v/f_\pi$.

Notice that, differently from the three-site model, in the two-site case there is no
reason to assume that the leading order term in the potential is cancelled
by a tuning among $y_L$ and $y_R$. The tuning of the potential can be totally
due to the counterterm which cancels the logarithmic divergence.
For this reason, in the following analysis we will not impose any relation between
the left and the right elementary--composite mixings.

In order to compute the coefficient $\beta$ at quartic order in $y_{L,R}$
we need to take into account an expansion of the logarithm in
eq.~(\ref{eq:potential_2-site}) at the quadratic order.
The value of the coefficient $\beta$ can be easily found analytically and is
given by
\begin{eqnarray}\label{eq:approx_B_2-site}
\beta &=& \frac{N_c}{8 \pi^2} \frac{({\widetilde m}_Q - {\widetilde m}_T)^2
y_L^2 y_R^2 f_\pi^4}{m_T^2 - m_{\widetilde T}^2} \log\left(\frac{m_T}{m_{\widetilde T}}\right)\nonumber\\
&& +\frac{N_c}{8 \pi^2}
\frac{({\widetilde m}_Q^2 - {\widetilde m}_T^2)^2 (y_L^2 - 2 y_R^2)^2 f_\pi^4
\left[-(m_T^2 - m_{\widetilde T}^2) + (m_T^2 + m_{\widetilde T}^2) \displaystyle
\log\left(\frac{m_T}{m_{\widetilde T}}\right)\right]
}{4 (m_T^2 - m_{\widetilde T}^2)^3}\,.
\end{eqnarray}
The term on the first line of the above expression
is analogous to the result found in the three-site case. On the other hand,
the second contribution is specific of the two-site model and is there because we did
not impose any relation between $y_L$ and $y_R$. The accidental factor of $4$ in the denominator
of the second contribution and some cancellations which happen in the expression between square
brackets make the second contribution smaller than the first one typically by
one order of magnitude. Notice, moreover, that the sign of the two contributions
are always the same. Thus the second contribution always determine a small increase of $\beta$
in absolute size.
Neglecting this second term we obtain a Higgs mass
\begin{equation}
m_H^2 = \frac{2 \beta}{f_\pi^2} \sin^2 (2 v/f_\pi)
\simeq \frac{N_c}{4 \pi^2} \frac{({\widetilde m}_Q - {\widetilde m}_T)^2
y_L^2 y_R^2 f_\pi^2}{m_T^2 - m_{\widetilde T}^2} \log\left(\frac{m_T}{m_{\widetilde T}}\right) \sin^2 (2 v/f_\pi)\,.
\label{mh22}
\end{equation}

As we did in the three-site model, we can rewrite the Higgs mass in terms of the
top mass. The approximate expression for the top mass was already found in
section~\ref{sec:genan} (eq.~(\ref{mt})) and is given by
\begin{equation}
m_t \simeq \frac{|\widetilde m_Q - \widetilde m_T|}{2 \sqrt{2}}
\frac{y_L y_R f_\pi^2}{m_T m_{\widetilde T}} \sin\left(\frac{2 v}{f_\pi}\right)\,.
\end{equation}
Making use of eq.~(\ref{mh22}) we find
\begin{equation}\label{eq:mHmt_2-site}
\frac{m_H}{m_t} \simeq \frac{\sqrt{2 N_c}}{\pi} \frac{m_{T} m_{{\widetilde T}}}{f_\pi}
\sqrt{\frac{\log\big(m_{T}/m_{{\widetilde T}}\big)}{m_{T}^2 - m_{{\widetilde T}}^2}}\,,
\end{equation}
which exactly coincides with the expression (\ref{eq:mHmt}) obtained in the three-site
model when the second level of resonances is heavy.

\subsection{Numerical results}

\begin{figure}[t]
\centering
\includegraphics[width=.455\textwidth]{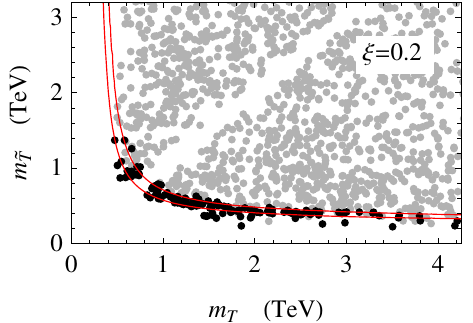}
\hspace{1.5em}
\includegraphics[width=.45\textwidth]{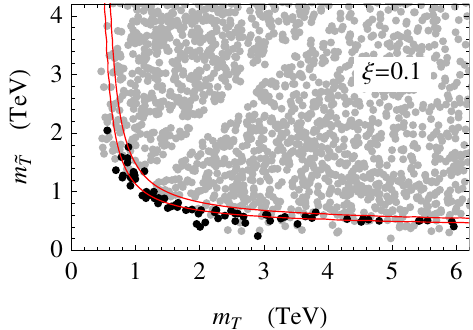}
\caption{Scatter plots of the masses of the $T$ and $\widetilde T$ resonances
for $\xi = 0.2$ (left panel) and $\xi = 0.1$ (right panel) in the two-site DCHM model. The black dots
denote the points for which $115\ {\rm GeV} \leq m_H \leq 130\ {\rm GeV}$, while
the gray dots have $m_H > 130\ {\rm GeV}$. The scans have been obtained by
varying all the composite sector masses in the range $[-8 f_\pi, 8 f_\pi]$ and keeping
the top mass fixed at the value $m_t = 150\ {\rm GeV}$.
The area between the solid red lines represents the range obtained by
applying the result in eq.~(\ref{eq:mHmt_2-site}) for
$115\ {\rm GeV} \leq m_H \leq 130\ {\rm GeV}$.}\label{fig:2-site_mBid_mSing}
\end{figure}

\begin{figure}[t]
\centering
\includegraphics[width=.455\textwidth]{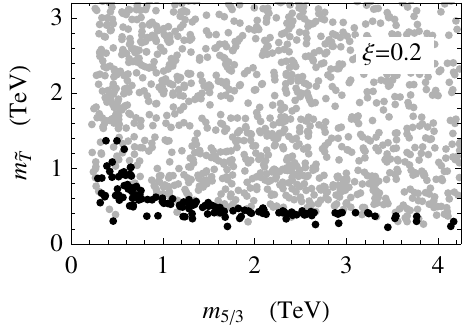}
\hspace{1.5em}
\includegraphics[width=.45\textwidth]{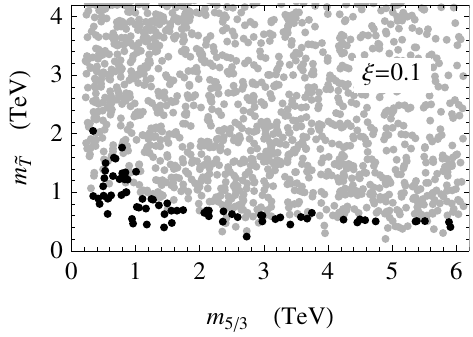}
\caption{Scatter plots of the masses of the exotic state of charge $5/3$
and of the $\widetilde T$ resonance for $\xi = 0.2$ (left panel) and $\xi = 0.1$ (right panel)
in the two-site DCHM model.
The black dots denote the points for which $115\ {\rm GeV} \leq m_H \leq 130\ {\rm GeV}$, while
the gray dots have $m_H > 130\ {\rm GeV}$. The scans have been obtained by
varying all the composite sector masses in the range $[-8 f_\pi, 8 f_\pi]$ and keeping
the top mass fixed at the value $m_t = 150\ {\rm GeV}$.}\label{fig:2-site_mEx_mSing}
\end{figure}

We can verify the validity of the relation in eq.~(\ref{eq:mHmt_2-site})
between the Higgs and the top partners masses by
performing a numerical scan on the parameter space of the two-site model.
However the computation of the Higgs effective
potential in the two-site case is not completely straightforward and requires
an ad hoc procedure to deal with the logarithmic divergence.
In particular, we can not directly integrate eq.~(\ref{eq:potential_2-site}) as in the $3$-site model.
The simplest way to proceed is to notice that eq.~(\ref{eq:potential_2-site})
can be rewriten in the standard Coleman--Weinberg form
\begin{equation}
V(h) = -\frac{2 N_c}{8 \pi^2} \int dp\ p^3 \log\left[
\prod_i (p^2 + m_i^2(h))\right]\,,
\label{CW}
\end{equation}
where the product is over all the $2/3$-charged fermionic states of our model. Actually, we could have derived eq.~(\ref{eq:potential_2-site})
starting from the Coleman--Weinberg expression in eq.~(\ref{CW}).
We can now regulate the integral with a hard momentum cutoff $\Lambda$ and
we obtain the standard formula
\begin{equation}\label{eq:ColWen}
V(h) = -\frac{N_c}{8 \pi^2}  \Lambda^2 \sum_i m_i^2(h)
- \frac{N_c}{16 \pi^2} \sum_i m_i^4(h) \left[\log \left(\frac{m_i^2(h)}{\Lambda^2}\right)-\frac12\right]\,.
\end{equation}

In the two-site model only a logarithmic divergence can appear in the Higgs potential,
and therefore the quadratically divergent term must be independent of the Higgs.
This is ensured by the condition
\begin{equation}
\sum_i m_i^2(h) = \sum_i m_i^2(h = 0) = {\rm const}.\,,
\end{equation}
which we can explicitly verify in our model. \footnote{
If, as in the $3$-site case, the Higgs potential was completely finite at one loop,
an analogous condition would hold for the
logarithmic term, {\it{i.e.}} $\sum_i m_i^4(h) = \sum_i m_i^4(h = 0) = {\rm const}.$}
The logarithmic divergence, as discussed above, must be proportional to $\sin^2{h/f_\pi}$ as in eq.~(\ref{eq:oper}).
Indeed in our $2$-site model one can verify explicitly that
$$
\sum_i m_i^4(h) \propto \sin(h/f_\pi^2) + {\rm const}\,.
$$
We can therefore, as anticipated, cancel the divergence by introducing a single counterterm
in the potential, proportional to $\sin^2{h/f_\pi}$.
This leaves only one free renormalization parameter which we can trade for a scale
$\mu$, the renormalized potential takes the form
\begin{equation}\label{Vren}
V(h) =
- \frac{N_c}{16 \pi^2} \sum_i m_i^4(h) \log \left(\frac{m_i^2(h)}{\mu^2}\right)\,.
\end{equation}
We will treat $\mu$ as a free parameter, the strategy of our scan will be to choose it,
once the other parameters are fixed, in order to fix the minimum
of the potential to the required value of $v/f_\pi$.

The result of the numerical scan is shown in fig.~\ref{fig:2-site_mBid_mSing}.
The black points correspond to configuration with
realistic Higgs mass and they lie approximately between the two solid
red lines which correspond to the bounds derived from eq.~(\ref{eq:mHmt_2-site}).
The small deviations come from the corrections due to the $(y_L^2 - 2 y_R^2)$
term in the expression for $\beta$ in eq.~(\ref{eq:approx_B_2-site}). As discussed
before, the effect of these corrections is to increase the Higgs mass, and therefore,
 keeping the Higgs mass fixed, to make the resonances lighter. In fig.~\ref{fig:2-site_mBid_mSing}
 we show the scatter plot of masses of the exotic charge $5/3$ state and of the $\widetilde T$.
 As in the three-site model the exotic state is lighter than the $T$, so that, in a large part of
the parameter space it is the lightest composite resonance.

\subsection{Modeling the effect of the heavy resonances}

By comparing the scatter plots obtained for the two-site model with the ones for the
three-site one, one can see that, although the relation between the Higgs
mass and the resonance masses is always reasonably well satisfied, significant deviations
can appear. In particular the region in which $m_T$ and $m_{\widetilde T}$ are comparable
shows larger deviations, while the asympthotic regions in which one of the resonances is
much lighter than the others have a smaller spread. The $2$-site model is therefore slightly too
restrictive, and also too ``pessimistic" in that it requires very low resonances. The effect of an
additional level of resonances, as the $3$-site model results show, can change the $2$-site
picture significantly.

However, the effect of the heavy resonances on the Higgs potential can be rather simply
mimicked in the two-site model by adding to the potential a new contribution to the
coefficient $\beta$ in eq.~(\ref{eq:pot_expansion_2-site}). The size of the
contributions coming from the heavy resonances can be estimated by symmetry
considerations and power counting. In our derivation we will respect the
general properties which characterize the heavy resonances in the three-site model.
First of all we assume that the source of ${\textrm{SO}}(5)$ breaking is in
common with the light states, so that the new operator must contain a factor
$({\widetilde m}_Q - {\widetilde m}_T)^2$. Morever we must introduce four powers of
the elementary--composite mixings as dictated by spurion analysis.
For simplicity we will write the contribution of the new operator to $\beta$
in the same form of the contribution coming from the light states. In particular
we choose the form of the most relevant term, the one on the first line
of eq.~(\ref{eq:approx_B_2-site}). Denoting by $M$ the mass of the heavy resonances
we write their contribution to the Higgs effective potential as
\begin{equation}\label{eq:heavypot}
\Delta V(h) = \frac{N_c}{8 \pi^2} \frac{({\widetilde m}_Q - {\widetilde m}_T)^2
y_L^2 y_R^2 f_\pi^4}{M^2} \sin^2(v/f_\pi) \cos^2(v/f_\pi)\,.
\end{equation}
Guided by the results of the three-site model, in which the heavy resonances tend
to lower the Higgs mass, we fix the sign for the corrections in order to
reproduce this effect.

\begin{figure}[t]
\centering
\includegraphics[width=.455\textwidth]{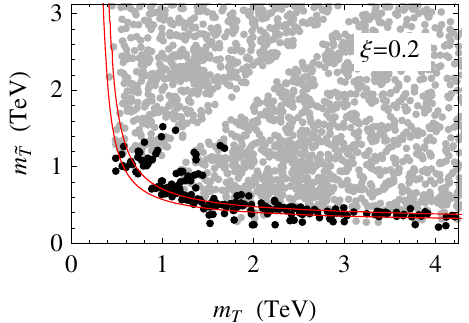}
\hspace{1.5em}
\includegraphics[width=.45\textwidth]{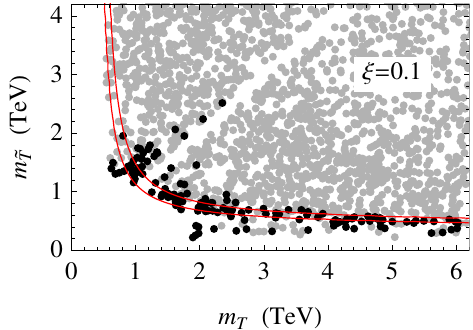}
\caption{Scatter plots of the masses of the $T$ and $\widetilde T$ resonances
for $\xi = 0.2$ (left panel) and $\xi = 0.1$ (right panel) in the two-site DCHM model
with the addition of the operator in eq.~(\ref{eq:heavypot}). The black dots
denote the points for which $115\ {\rm GeV} \leq m_H \leq 130\ {\rm GeV}$, while
the gray dots have $m_H > 130\ {\rm GeV}$. The scans have been obtained by
varying all the composite sector masses in the range $[-8 f_\pi, 8 f_\pi]$ and keeping
the top mass fixed at the value $m_t = 150\ {\rm GeV}$. The mass of the
heavy resonances has been chosen to be at least $50\%$ higher than the one of
all the light states. The area between the solid red lines represents the range obtained by
applying the result in eq.~(\ref{eq:mHmt_2-site}) for
$115\ {\rm GeV} \leq m_H \leq 130\ {\rm GeV}$.}\label{fig:2-site_mBid_mSing_mod}
\end{figure}

\begin{figure}[t]
\centering
\includegraphics[width=.455\textwidth]{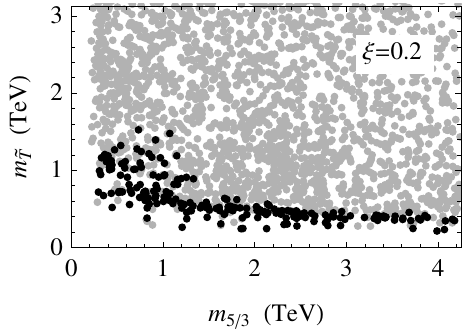}
\hspace{1.5em}
\includegraphics[width=.45\textwidth]{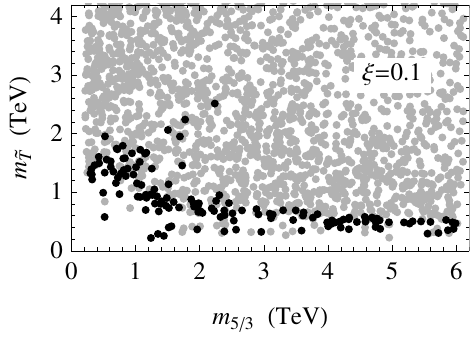}
\caption{Scatter plots of the masses of the exotic state of charge $5/3$
and of the $\widetilde T$ resonance for $\xi = 0.2$ (left panel) and $\xi = 0.1$ (right panel)
in the two-site DCHM model with the addition of the operator in eq.~(\ref{eq:heavypot}).
The black dots denote the points for which $115\ {\rm GeV} \leq m_H \leq 130\ {\rm GeV}$, while
the gray dots have $m_H > 130\ {\rm GeV}$. The scans have been obtained by
varying all the composite sector masses in the range $[-8 f_\pi, 8 f_\pi]$ and keeping
the top mass fixed at the value $m_t = 150\ {\rm GeV}$.  The mass of the
heavy resonances has been chosen to be at least $50\%$ higher than the one of
all the light states.}\label{fig:2-site_mEx_mSing_mod}
\end{figure}

The numerical results of a scan including the effect of the operator in
eq.~(\ref{eq:heavypot}) are shown in fig.~\ref{fig:2-site_mBid_mSing_mod}.
In the scan we assume that the mass of the heavy resonances is at least
$50\%$ higher than the masses of all the light resonances.
One can see that the plots show a good qualitative and quantitative agreement
with the ones obtained in the three-site model (see fig.~\ref{fig:mBid_mSing_DCHM3}).
In particular the plots show an agreement with the relation in eq.~(\ref{eq:mHmt_2-site})
in the asymptotic regions in which one state is much lighter than the others.
Larger deviations are present when all the state have comparable masses.
This effect can be simply understood by comparing the form of the leading
contributions to $\beta$ in eq.~(\ref{eq:approx_B_2-site}) (the ones on the first line)
and the form of the contributions of the operator representing the heavy resonances
in eq.~(\ref{eq:heavypot}). When a high hierarchy between $m_T$ and $m_{\widetilde T}$
is present, the logarithm appearing in eq.~(\ref{eq:approx_B_2-site}) enhances
the light states contributions to the Higgs mass, thus making the
heavy resonances corrections negligible. On the other hand, when
$m_T \sim m_{\widetilde T}$, the light states contribution are somewhat reduced
and the heavy states can give a sizable correction to $\beta$.

Finally in fig.~\ref{fig:2-site_mEx_mSing_mod} we show the scatter plot for
the masses of the exotic charge $5/3$ state and of the ${\widetilde T}$ state.
Again a good agreement with the results for the three-site model
in fig.~\ref{fig:mEx_mSing_DCHM3} is found.

\section{Bounds on the top partners}\label{sec:exp_excl}

The top partners are generically so light, often below the TeV, than the present experimental results can
already place some non-trivial bounds on their mass. In this section we will present a simple discussion of
the available constraints; our aim will not be to perform a comprehensive study
of all the bound coming from the existing experimental data, but instead to
focus on some simple and universal searches whose results are
approximately valid independently of the specific model and of the corner of the parameter
space we consider.

In particular we will restrict our analysis to the lightest resonance
which comes from the composite sector and we will only consider
pair production processes in which, due to the universal QCD couplings,
the production cross section depends exclusively on the mass of the resonance.
The bounds we will derive are thus quite robust and apply to generic
composite models. Notice however that, in a large region of the parameter space,
single production processes, as well as the presence of other relatively
light resonances, can give an enhancement of the signal
in the channels considered in the present analysis. In this case
the bounds on the masses of the resonances can also become tighter.
Taking into account these effects is however beyond the scope of the present paper.

\begin{figure}[t]
\centering
\includegraphics[width=.45\textwidth]{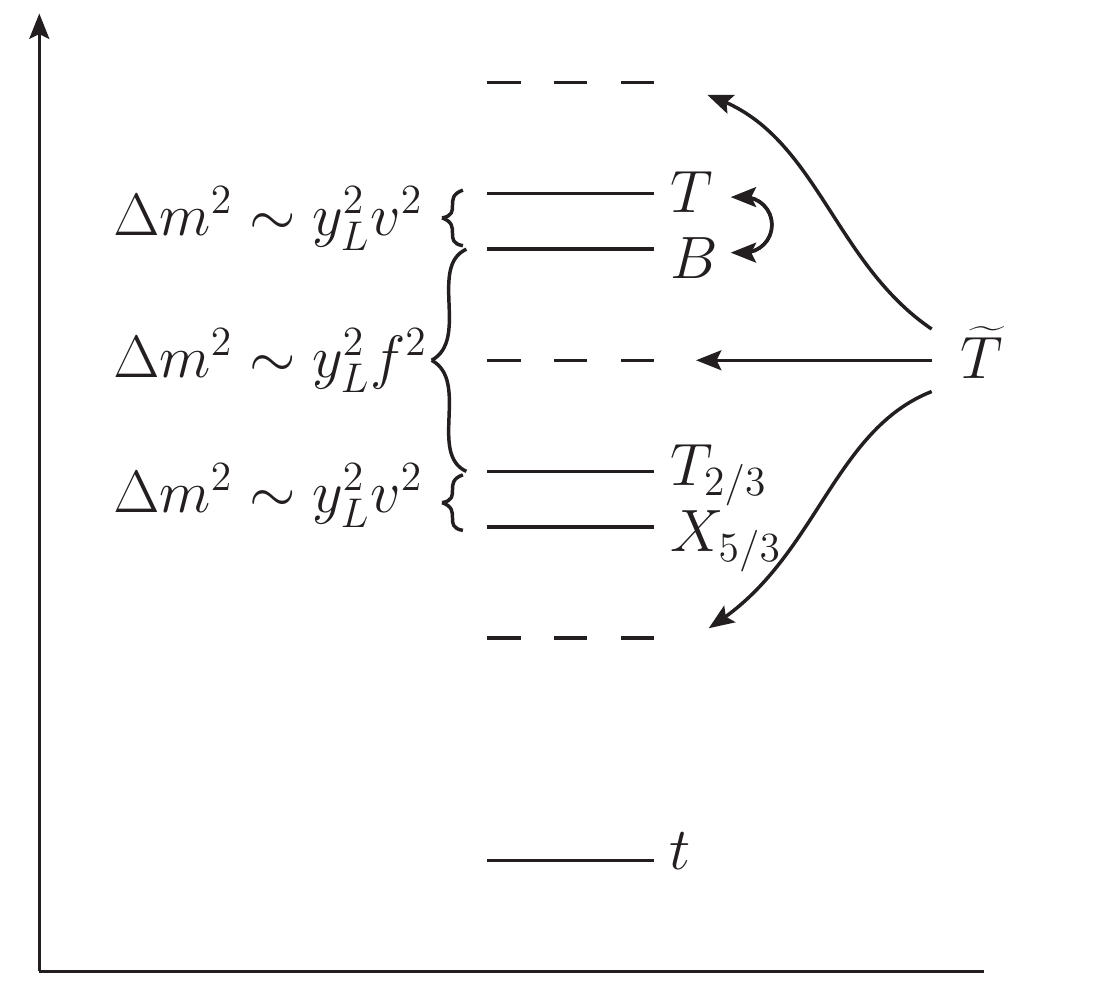}
\caption{Schematic structure of the spectrum of the lightest multiplet of
resonances.}\label{fig:Mass_spectrum}
\end{figure}

Before discussing the details of our analysis, it is useful to briefly
describe the general structure of the spectrum of the first level of
fermionic resonances. These states, as schematically shown in fig.~\ref{fig:Mass_spectrum},
are approximately organized in $SU(2)_L$ multiplets
\begin{equation}
Q =
\left(
\begin{array}{cc}
T \\
B
\end{array}
\right)\,,
\qquad
X =
\left(
\begin{array}{cc}
X_{5/3} \\
T_{2/3}
\end{array}
\right)\,,
\qquad
\widetilde{T}\,.
\end{equation}
The splitting between the two doublets arises from the mixing of the composite
fermions with the elementary states and its size is of order $\Delta m^2 \sim y_L^2 f^2$.
Notice that only the $Q$ doublet is mixed to the elementary fermions, thus
it is always heavier than the $X$ doublet.
On the other hand, the mass of the $\widetilde T$ singlet has no relation to the
ones of the two doublets.

After the breaking of the electroweak symmetry the fermions acquire mass corrections
giving rise to a small splitting inside the doublets. Due to the Goldstone nature of the Higgs,
the effects of EWSB can only arise if the Goldstone symmetry is broken, that is they
must be mediated by the elementary--composite mixings. The mass splitting inside
the doublets are thus of order $y_{L,R}^2 v^2$, and are typically suppressed
by a factor $(v/f)^2$ with respect to the mass gap between the two doublets.
For all the relevant configurations the lightest state of the $X$ doublet is
the exotic fermion with charge $5/3$, the $X_{5/3}$. The ordering of the states in the $Q$
multiplet instead is not fixed and depends on the specific point in the parameter
space we choose.

As we mentioned before, in our analysis we will only consider the lightest
fermionic resonance, which is always given by the exotic state $X_{5/3}$ or by the
singlet $\widetilde T$. We will discuss these two cases separately in the following
subsections.

\subsection{Bounds on the exotic charge $5/3$ state}

As a first case we will consider the configurations in which the exotic state $X_{5/3}$
is the lightest new resonance. A search for an exotic state of this type has been
performed by the CDF collaboration~\cite{Aaltonen:2009nr}. This analysis focuses
on the case in which the exotic state is associate with a charge $-1/3$ fermion, the $B$,
with the same mass. Moreover it is assumed that these two states always decay in $t W^{\pm}$.
The considered channel is pair production of the new resonances, which then give rise
to a signal in events with two same-sign leptons. With an integrated luminosity
of $2.7\ {\rm fb}^{-1}$, masses of the new states $m_{5/3} = m_B < 365\ {\rm GeV}$
are excluded at $95\%$ confidence level.

Dedicated searches for charge $5/3$ states are currently not available for the LHC data.
Some interesting bounds on the mass of the exotic states can however be derived by adapting
the existing searches of new bottom-like resonances. The analyses we can use for our aim are
the ones in which the bottom-like resonance $B$ is pair produced and decays in top:
$B\overline B \rightarrow W^- t W^+ \overline t$. The same final state is obviously obtained
also in a process in which a pair of exotic states $X_{5/3}$ are produced, which then decay
to SM particles: $X_{5/3}\overline X_{5/3} \rightarrow W^+ t W^- \overline t$.

The strongest exclusion bound on new bottom-like quarks decaying in tops is the one
obtained by the CMS Collaboration~\cite{CMS:BtoWt}, which sets a lower bound
$m_B > 611\ {\rm GeV}$ at $95\%$ confidence level assuming $BR(B \rightarrow W^- t) = 1$.
This analysis is performed by condidering final states with a pair of same-sign leptons
or with three leptons. \footnote{Less stringent bounds have been obtained
by the ATLAS Collaboration, whose public analysis, for same-sign dilepton final states,
reports a bound $m_B > 450\ {\rm GeV}$~\cite{Aad:2012bb}. An analysis with final states
with a single lepton and multiple jets has also been published by the ATLAS
Collaboration, in which a bound $m_B > 480\ {\rm GeV}$ is reported~\cite{Aad:2012us}.}
To translate this result into an exclusion bound for the exotic $X_{5/3}$ resonance,
we need to take into account possible differences in the efficiencies for
the cuts used in the analysis. These differences can arise from finite-width effects
and from the different kinematic distribution of the final states of the two
processes. In particular in the $X_{5/3}\overline X_{5/3}$ process the two same-sign
leptons come from the decay of the same heavy particle, while in
the $B\overline B$ case they come from different heavy legs.
The structure of the cuts used in the analysis, however, is rather symmetric with respect
to the leptons, so we expect the efficiencies to be reasonably close.
To determine the variation of the cut acceptances we simulated the signal using
our implementation of the three- and two-site models in {\sc MadGraph} 5.
We found that the deviations from the $B\overline B$ process
are always negligible (below $5\%$).

The production cross sections for $X_{5/3}\overline X_{5/3}$ and $B\overline B$
are equal, given that the two processes are induced by QCD. Moreover, being the $X_{5/3}$
the lightest resonance, it can only decay to the SM, so that $BR(X_{5/3} \rightarrow W^+ t) = 1$.
This means that we can directly reinterpret the exclusion bound on the bottom-like
resonances as a lower bound on the mass of the exotic state $X_{5/3}$: $m_{5/3} > 611\ {\rm GeV}$.

Notice that the presence of other resonances with a mass relatively close to the
$X_{5/3}$, and in particular a light $B$, can sizably enhance the signal used in
the previous analysis, thus leading to stronger bounds. The analysis of this effect is
however beyond the scope of the present paper.

\subsection{Bounds on the $\widetilde T$}

We now focus on the case in which the lightest resonance is given by the
charge $2/3$ state $\widetilde T$. For a state of this kind exclusion analysis
have been performed for the available LHC data. At present the strongest
bounds are the ones obtained by the CMS Collaboration. They considered two
possible scenarios in which a top-like resonance decays with $100\%$ branching ratio
either in $tZ$, yielding a bound $m_{t'} > 475\ {\rm GeV}$~\cite{CMS:TtoZt},
or in $bW^+$, with bounds $m_{t'} > 560\ {\rm GeV}$~\cite{CMS:TtoWb}
and $m_{t'} > 552\ {\rm GeV}$~\cite{CMS:TtoWb_2} depending on the specific
final states considered. \footnote{A search for a $t'$ states decaying in $b W^+$
has been performed also by the ATLAS Collaboration, which derives a
bound $m_{t'} > 404\ {\rm GeV}$~\cite{Aad:2012xc}.}

\begin{figure}[t]
\centering
\includegraphics[width=.45\textwidth]{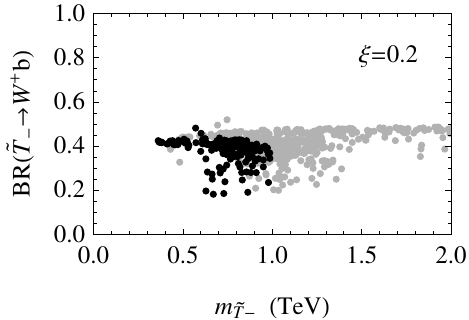}
\hspace{1.5em}
\includegraphics[width=.45\textwidth]{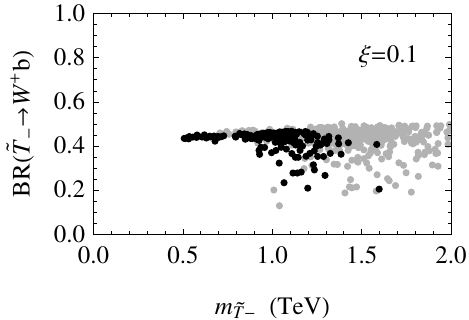}
\caption{Scatter plot of the branching ratios of the lightest $\widetilde T$
resonance into $W^+ b$ for the three-site DCHM model
with $\xi = 0.2$ (left panel) and $\xi = 0.1$ (right panel).
In all the points shown in the plot the $\widetilde T$ state has been
required to be the lightest composite resonance. The black dots denote
the points for which $115\ {\rm GeV} \leq m_H \leq 130\ {\rm GeV}$, while
the gray dots have $m_H > 130\ {\rm GeV}$. The scans have been obtained
by varying all the composite sector masses in the range $[-8 f, 8f]$ ane
keeping the top mass fixed to the value $m_t = 150\ {\rm GeV}$.}\label{fig:BR_Ttilda_Wb_3s}
\end{figure}

These bounds can not be directly translated into bounds on the $\widetilde T$ resonance
in our model, due to the different branching ratios of the resonance into
SM particles. In particular three channels are relevant $\widetilde T \rightarrow b W^+$,
$\widetilde T \rightarrow t Z$ and $\widetilde T \rightarrow t h$. The branching
ratios for these three channels are all comparable, hence a process in which
two $\widetilde T$ resonances are produced, which then decay in the same channel,
has a cross section which is usually an order of magnitude smaller than the
total pair production cross section. From a scan on the parameter space of the
explicit models, see fig.~\ref{fig:BR_Ttilda_Wb_3s}, we find that typically the $W$ and Higgs
channels dominate,
$BR(\widetilde T \rightarrow b W^+) \sim BR(\widetilde T \rightarrow t h) \sim 0.4$,
while the $Z$ channel is slightly suppressed, $BR(\widetilde T \rightarrow t Z) \sim 0.2$.

To find an exclusion bound on the $\widetilde T$ resonance we adopt the
simple and conservative approach of just rescaling the cross section of each channel
considered in the experimental analysis by the typical branching ratios
predicted by our models for a low resonance mass. Of course, a more refined procedure,
would need to take into account possible enhancements of the signal coming
from the other decay channels. For instance, in the search of a top-like
resonance decaying in $Z t$~\cite{CMS:TtoZt} the masses of the resonances
are not reconstructed and only a mild cut is put to reconstruct one of the $Z$'s.
In this case a sizable part of events in which one or even both $\widetilde T$
resonances do not decay in $Z t$ could pass the selection cuts and significantly
enhance the signal, thus tightening the exclusion bounds.

Following our simple approach we find that only one of the searches gives a significant bound,
namely the one exploiting the channel $\widetilde T \overline {\widetilde T} \rightarrow b W^+ \overline b W^-
\rightarrow b l^+ \nu \overline b l^- \overline \nu$~\cite{CMS:TtoWb_2}.
To obtain the bound we performed the analysis using for the relevant branching ratio
the value $BR(\widetilde T \rightarrow b W^+) = 0.4$. As can be seen by the
scatter plot in fig.~\ref{fig:BR_Ttilda_Wb_3s}, this value represents the
actual branching ratio for a light $\widetilde T$ in the three-site
DCHM model, quite independently of the value of $v/f_\pi$.
With this procedure we infer a lower bound $m_{\widetilde T} > 370\ {\rm GeV}$
on the mass of the $\widetilde T$ resonance at $95\%$ confidence level.

\subsection{Exclusion bounds in the DCHM$_3$}

\begin{figure}[t]
\centering
\includegraphics[width=.45\textwidth]{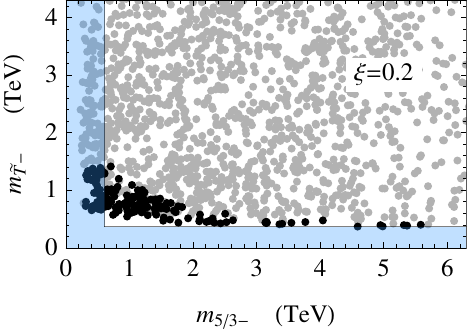}
\hspace{1.5em}
\includegraphics[width=.45\textwidth]{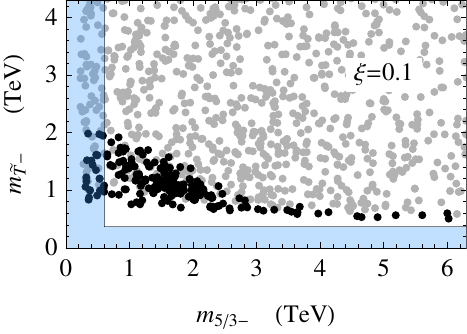}
\caption{Scatter plot of the masses of the lightest exotic state of charge $5/3$
and of the lightest $\widetilde T$ resonance for the three-site DCHM model
with $\xi = 0.2$ (left panel) and $\xi = 0.1$ (right panel).
The shaded region corresponds to the points excluded by our analysis, which gives the bounds
$m_{5/3} > 611\ {\rm GeV}$ and $m_{\widetilde T} > 370\ {\rm GeV}$.
The black dots denote the points for which $115\ {\rm GeV} \leq m_H \leq 130\ {\rm GeV}$, while
the gray dots have $m_H > 130\ {\rm GeV}$.}\label{fig:3-site_Exclusion}
\end{figure}

To appreciate the impact of the previously derived bounds in the explicit models
we show in fig.~\ref{fig:3-site_Exclusion} the exclusion regions superimposed on the
scatter plots for the masses of the $X_{5/3}$ and $\widetilde T$ resonances for the
three-site DCHM model.

The bound on the exotic state with chagre $5/3$ is already strong enough to exclude
a sizable portion of the parameter space with realistic Higgs mass. Of course,
the bound has a greater impact on the configurations with larger $\xi$, which
predict lighter resonances. Nevertheless even in the case of a relatively small
$v/f_\pi$, namely $\xi = 0.1$, the exclusion bound on the exotic resonance puts
non-trivial constraints.

The situation is different for the cases in which the lightest new state is
the singlet $\widetilde T$. The bounds obtained in our analysis can only
exlude a limited number of configurations at $\xi$ relatively large. In particular
for $\xi = 0.2$ realistic configurations start to be excluded
only in the asymptotic region with a light $\widetilde T$. On the other hand,
for $\xi = 0.1$ the mass of the $\widetilde T$ resonances is always above the
current bounds.

\section{Conclusions and outlook}

In this paper we explored the relation which, in a broad class of composite Higgs
models, links a light Higgs boson to the presence of light resonances coming
from the new strong sector of the theory. The class of models we focused on are the ones
based on the symmetry pattern ${\textrm{SO(5)/SO(4)}}$ and in which the
fermionic resonances come in the fundamental representation of ${\textrm{SO(5)}}$.
As a first step, we analyzed from a
general point of view the mechanism which generates the correlation of the Higgs mass
with the top partners. We found that the connection has a simple qualitative explanation
in terms of the structure of partial fermion compositeness which is realized
in the composite Higgs scenario.
The point is that the presence of light partners would tend to increase the fraction of top quark compositeness,
thus increasing its mass. Keeping the latter fixed requires that the elementary--composite mixings,
must be decreased in order to compensate.
But the mixings also control the Higgs potential, and their decrease lowers the the Higgs quartic
coupling and consequently the Higgs mass.

Through a detailed analysis performed in a general effective parametrization of the
composite Higgs set-up, we found that the Higgs mass scales linearly
with the mass of the lightest partner. This model-independent result
is encoded in the simple relation of eq.~(\ref{eq:mhgen}). The relevant partners
are those which are most strongly mixed with the elementary $t_L$ and $t_R$,
namely the $T$ and the $\widetilde T$ resonances.

From a quantitative point of view, assuming only a moderate
degree of tuning between the Higgs VEV and the Goldstone decay constant $f_\pi$,
we found that a Higgs mass in the current LHC preferred region $m_h \simeq 120\ {\rm GeV}$ requires
at least one top partner with a mass of the order or below the ${\rm TeV}$.
This result strengthens the common assertion that light top partners are an
essential feature of the composite Higgs scenario. Moreover it shows that
the lightest of such states are well within the reach of the LHC and
they constitute one of the most important probes of the composite Higgs paradigm.

To confirm the validity of our general results we analyzed in detail the relation
between the Higgs and the resonance masses in two explicit models. The first model
we considered is the three-site Discrete Composite Higgs Model, which gives
a simple but complete realization of the composite Higgs framework. Good agreement with
the general result is found (see eq.~(\ref{eq:mHmt})). The study of this explicit model
allowed us to quantify the effects of the higher levels of resonances
coming from the composite sector. Their contribution
has been determined analytically (see eq.~(\ref{eq:exact_B})) and checked numerically
by a scan on the parameter space of the model (see figs.~\ref{fig:mBid_mSing_DCHM3}
and \ref{fig:mEx_mSing_DCHM3}). We found that in the asymptotic regions in which
there is a large mass hierarchy between the resonances of the first level,
the effect of the heavier states is very small. On the contrary, when the masses are comparable the
corrections can be sizable and can affect the Higgs mass by as much as $50\%$. These corrections, however,
do not spoil the agreement with the qualitative picture obtained by the general analysis.

As a second explicit model, we considered an even simpler and more minimal implementation of the
composite Higgs idea, the two-site Discrete Composite Higgs Model. This model contains only one
level of composite resonances, thus it only retains the amount of information
relevant for the present collider experiments. The price we have to pay in
adopting this minimal description is a partial loss of predictivity: in the two-site
model the Higgs potential is no more finite at one loop. Nevertheless only
one counterterm is needed to regulate the divergence, thus one condition
is enough to fix the renormalization ambiguity. We can therefore chose the value of $v/f_\pi$
as a renormalization condition and retain the Higgs mass as a \emph{calculable} quantity.

We also showed how the effect of the higher resonance levels on the Higgs potential can be
efficiently modeled by introducing a suitable extra contribution.
With this modification, the spectrum of the top partners in correlation with the Higgs mass
is completely analogous to the one found in the three-site case.

Generically, our result should apply to any explicit realization of the composite Higgs idea, and in
particular to the popular 5d holographic models. This is partially confirmed by the numerical results
of Ref.~\cite{Contino:2006qr}, which indeed show an approximately linear
correlation between the mass of the Higgs and the one of the lightest $T$ state. From what can
be seen in the plots, moreover, the agreement with our general formula seems good also at a quantitative
level. It might be worth checking the agreement in more detail.

Another aspect which could be worth investigating is how much our results depend on the
choice of the fermion representations.
In the set-ups considered so far in the literature, in which the fermions
are in the spinorial or in the adjoint representation of ${\textrm{SO(5)}}$ there will be
no qualitative difference with respect to the case of the fundamental we have considered
in the present paper. The general analysis of section~2 will apply in the same way because
also in these alternative scenarios only one invariant operator contributes to the Higgs potential
at the leading order in the elementary--composite mixing. This implies that the leading order must be
canceled and the Higgs mass squared scales as $y^4$ (rather than $y^2$) like in eq.~(\ref{mh}).
Also the estimate of the top mass
will remain the same and therefore the final result of eq.~(\ref{eq:mhgen}) will be parametrically
unchanged. At the quantitative level, however, the relation among the Higgs and the resonance
masses could be modified at order one, due to possibly different group theory factors.
It would be interesting to assess this point.

With other choices of the fermion representations, in which two or more invariants appear at
order $y^2$, our conclusions could instead change qualitatively. A model of this kind could for instance
be obtained by embedding the fermions in the adjoint but relaxing the requirement
of left--right symmetry, \footnote{Introducing a breaking of the left--right symmetry would of course
lead to potentially large deviations in the $Zb_L\overline b_L$ coupling, which
could invalidate the model or lead to additional tuning.} or by considering higher and possibly
reducible representations.

The models with more invariants are particularly interesting because they do not suffer of the
enhanced (or ``double'') tuning which we described in section~2 (see eq.~(\ref{tunn})).
This indeed originates only in the case of a single
invariant because of the ``preliminary'' cancellation of the ${\cal O}(y^2)$ term which is needed
to make it of the same order of the subleading ${\cal O}(y^4)$ contributions.
Models in which the double tuning is not present, however, should face another
possible problem. Given that the potential, including the quartic Higgs coupling,
now arises at ${\cal O}(y^2)$, the mass of the Higgs is expected to be a factor
$g_*/y$ \emph{larger} than in the models with the preliminary cancellation.
The situation would therefore be worse than the one considered in the present
paper, so we expect that either we will find a too large Higgs mass or too light top partners.

The last step of our work has been to derive
some non trivial constraints on the resonance masses from the current LHC data.
We performed a simple analysis in which only the lightest resonances have been
considered and the contributions from heavier states have been altogether neglected.
The lightest composite states are either the exotic partner of charge $5/3$, the $X_{5/3}$,
or the ${\textrm{SO(4)}}$ singlet top-like state $\widetilde T$.
No dedicated LHC analysis exists for states of charge $5/3$, however
the existing searches for pair-produced new bottom-like quarks can be used to derive some
exclusion bounds also for the exotic resonance. Our simulations show that the current
exclusion can be translated into a bound $m_{5/3} > 611\ {\rm GeV}$. This bound is already enough
to exclude a non-negligible portion of the configurations with a realistic Higgs mass.

The situation is different for the cases in which the lightest state is the $\widetilde T$.
LHC searches for pair-produced charge $2/3$ resonances are available,
however the bounds given in the experimental analyses can not be naively applied to the
composite models, due to the fact that $100\%$ branching ratios to specific channels
are assumed. Taking into account the branching ratios predicted by our explicit models,
a lower bound $m_{\widetilde T} > 370\ {\rm GeV}$ is obtained. This bound is still
too weak to put significant constraints on the model. However it is not far from
the masses obtained in realistic configurations and can already exclude a small
region of the parameter space in the case $\xi = 0.2$.

The simple analysis we used to derive the exclusion bounds leads to some
robust and conservative constraints on the masses of the composite resonances.
The assumption of having only one relevant light state and to consider
only pair production is however a drastic simplification which often neglects
 important contributions. In particular some of the heavier states can be close
enough to the lightest resonance to give a sizable contribution to the relevant search channels.
Moreover single production processes can be relevant in specific regions
of the parameter space. It is plausible that the inclusion of these effects
could considerably strengthen the bounds we derived in this paper.

A complete study of the exclusion bound, including the effects of the heavier
resonances and of single production, is not completely straightforward.
An analysis of this kind should take into account possible decay chains
from the heavier resonances as well as non-universal production cross
sections due to single production channels. To perform this task it seems unavoidable
to employ a concrete model. We already produced
complete {\sc MadGraph} 5 cards implementing the two-site and the three-site models.
In particular the two-site model contains only three free parameters and
one could imagine performing a complete scan.
These cards were used in the present paper to adapt the experimental
analysis and derive the bounds on the exotic state $X_{5/3}$.
We leave for future work the complete study of the constraints coming from the
experimental data.

\section*{Acknowledgments}

We thank R.~Contino, F.~Goertz, M.~Redi, R.~Rattazzi and M.~Serone for discussions.
We also thank D.~Marzocca and M.~Serone \cite{Serone}
and M.~Redi and A.~Tesi \cite{Redi} for sharing with us before publication
the results of their works.
The work of A.~W. and O.~M was supported in part by the European Programme Unification
in the LHC Era, contract PITN-GA-2009-237920 (UNILHC) and by the ERC Advanced Grant no.267985
Electroweak Symmetry Breaking, Flavour and Dark Matter: One Solution for Three Mysteries (DaMeSyFla).

\end{document}